\documentclass[fleqn,usenatbib]{mnras}

\usepackage{newtxtext}
\usepackage[varvw]{newtxmath} 

\usepackage[T1]{fontenc}

\DeclareRobustCommand{\VAN}[3]{#2}
\let\VANthebibliography\thebibliography
\def\thebibliography{\DeclareRobustCommand{\VAN}[3]{##3}\VANthebibliography}

\usepackage{graphicx}	
\usepackage{amsmath}	
\usepackage{hyperref}   
\usepackage[rightcaption]{sidecap}
\sidecaptionvpos{figure}{c}
\usepackage[labelfont=bf]{caption} 
\usepackage{lscape}
\usepackage{array}  
\usepackage{multirow}  
\usepackage{caption}
\usepackage{subcaption}

\newcommand{\avir}{\alpha_\mathrm{vir}}  
\newcommand{\fmmc}{f_\mathrm{MMC}}
\newcommand{\fc}{f_\mathrm{C}}
\newcommand{\firb}{f_\mathrm{IRB}}
\newcommand{\qmmc}{q_\mathrm{MMC}}

\newcommand{\mmmc}{M_\mathrm{MMC}}
\newcommand{\mpc}{M_\mathrm{1pc}}
\newcommand{\sigpc}{\Sigma_\mathrm{1pc}}

\newcommand{\ncomp}{\bar{N}_\mathrm{comp}}
\newcommand{\sigcomp}{\bar{\sigma}_\mathrm{comp}}
\newcommand{\sigline}{\bar{\sigma}_\mathrm{line}}
\newcommand{\sigtotal}{\bar{\sigma}_\mathrm{total}}
\newcommand{\vcen}{v_\mathrm{cen}}

\newcommand{\kms}{km$\,$s$^{-1}$}
\newcommand{\Lsun}{$\mathrm{L}_\odot$}
\newcommand{\mic}{$\upmu$m}
\newcommand{\mjybeam}{mJy$\,$beam$^{-1}$}
\newcommand{\Msun}{$\mathrm{M}_\odot$}
\newcommand{\pcmmm}{\,cm$^{-3}$}	

\newcommand{\ceo}{C$^{18}$O}
\newcommand{\hcop}{HCO$^+$}
\newcommand{\nthp}{N$_2$H$^+$}

\newcommand{\mwydyn}{{\sc mwydyn}}

\newcolumntype{L}[1]{>{\raggedright\let\newline\\\arraybackslash\hspace{0pt}}b{#1}}   

\title[The dynamic centres of IRDCs]{The dynamic centres of infrared-dark clouds and the formation of cores}

\author[A. J. Rigby et al.]{
Andrew J. Rigby$^{1,2}$\thanks{E-mail: a.j.rigby@leeds.ac.uk},
Nicolas Peretto$^2$,
Michael Anderson$^2$,
Sarah E. Ragan$^2$,
Felix D. Priestley$^2$,
\newauthor 
Gary A. Fuller$^{3,4}$,
Mark A. Thompson$^1$,
Alessio Traficante$^5$,
Elizabeth J. Watkins$^{3,6}$,
\newauthor and Gwenllian M. Williams$^{1,7}$.
\\
$^{1}$School of Physics and Astronomy, University of Leeds, Leeds LS2 9JT, UK\\
$^{2}$Cardiff Hub for Astrophysics Research \& Technology, School of Physics \& Astronomy, Cardiff University, Queen's Buildings, The Parade, Cardiff, CF24 3AA, \\UK\\
$^{3}$Jodrell Bank Centre for Astrophysics, School of Physics and Astronomy, University of Manchester, Oxford Road, Manchester, M13 9PL, UK\\
$^{4}$Physikalisches Institut, University of Cologne, Z{\"u}lpicher Str. 77, 50937 K{\"o}ln, Germany\\
$^{5}$IAPS-INAF, Via Fosso del Cavaliere, 100, I-00133, Rome, Italy\\
$^{6}$Astronomisches Rechen-Institut, Zentrum f{\"u}r Astronomie der Universit{\"a}t Heidelberg, M{\"o}nchhofstra{\ss}e 12-14, D-69120 Heidelberg, Germany\\
$^{7}$Department of Physics, Aberystwyth University, Ceredigion, Cymru, SY23 3BZ, UK
}

\date{Accepted XXX. Received YYY; in original form ZZZ}

\pubyear{2024}

\begin{document}
\label{firstpage}
\pagerange{\pageref{firstpage}--\pageref{lastpage}}
\maketitle

\begin{abstract} 
High-mass stars have an enormous influence on the evolution of the interstellar medium in galaxies, so it is important that we understand how they form. We examine the central clumps within a sample of seven infrared-dark clouds (IRDCs) with a range of masses and morphologies. We use 1\,pc-scale observations from the Northern Extended Millimeter Array (NOEMA) and the \emph{IRAM} 30-m telescope to trace dense cores with 2.8\,mm continuum, and gas kinematics in \ceo, \hcop, HNC, and \nthp\ ($J$=1--0). We supplement our continuum sample with six IRDCs observed at 2.9\,mm with the Atacama Large Millimeter/submillimeter Array (ALMA), and examine the relationships between core- and clump-scale properties. We have developed a fully-automated multiple-velocity component hyperfine line-fitting code called \mwydyn\ which we employ to trace the dense gas kinematics in N$_2$H$^+$ (1--0), revealing highly complex and dynamic clump interiors. We find that parsec-scale clump mass is the most important factor driving the evolution; more massive clumps are able to concentrate more mass into their most massive cores -- with a log-normally distributed efficiency of around 9\% -- in addition to containing the most dynamic gas. Distributions of linewidths within the most massive cores are similar to the ambient gas, suggesting that they are not dynamically decoupled, but are similarly chaotic. A number of studies have previously suggested that clumps are globally collapsing; in such a scenario, the observed kinematics of clump centres would be the direct result of gravity-driven mass inflows that become ever more complex as the clumps evolve, which in turn leads to the chaotic mass growth of their core populations. 
\end{abstract} 




\section{Introduction} \label{sec:introduction}

The stellar populations of the many billions of galaxies in the Universe are dominated, in absolute number, by low-mass stars ($m_* \lesssim 2\,\mathrm{M}_\odot$), while the much rarer high-mass stars ($m_* > 8\,\mathrm{M}_\odot$) have an enormous influence on the interstellar medium (ISM) and chemical evolution of galaxies. The relative abundance of low- through to high--mass stars as they enter the main sequence is given by the stellar initial mass function (IMF), and our understanding of the evolution of galaxies depends critically upon our understanding of how the IMF arises.

For low-mass stars, observations of nearby star-forming regions have revealed that the gravitational fragmentation of filaments with a (super-)critical mass-per-unit-length appears to play a crucial role in determining the masses of 0.1\,pc-scale cores \citep{Andre+14}. These cores are thought to be the precursors of individual stellar systems, and the distribution of core masses -- the core mass function (CMF) -- is found in these regions to closely resemble the IMF, leading to speculation that the latter is inherited directly from the former \citep[e.g][]{Andre+10, Polychroni+13, Andre+14, Konyves+20, Ladjelate+20}. However, for high-mass stars, the picture is more complicated \citep[see][for a recent review]{Motte+18a}. High-mass star-forming regions are inherently more difficult to observe as a consequence of both their short lifetimes ($\sim$0.5--2\,Myr, \citealt{Battersby+17, Sabatini+21}), and their relatively low rate of occurrence. The relatively small number of observations of high-mass star-forming regions that are able to robustly measure the CMF find them to deviate significantly from the IMF, with top-heavy distributions \citep[e.g.][]{Motte+18, Kong19}, and fragmentation studies \citep[e.g.][]{Bontemps+10, Sanhueza+17} routinely demonstrate that Jeans-like fragmentation is unable to produce cores that could be the progenitors of high-mass stars, indicating a different formation pathway.

One of the most massive protostellar cores that has been observed in the Galaxy, SDC335-MM1 \citep[$\sim$500$\,$\Msun;][]{Peretto+13}, is located at the centre of a so-called `hub-filament' system \citep[HFS;][]{Myers09}: a focal point of converging filaments of dense gas. A growing sample of HFSs has since been studied in the literature, which often find massive cores at the centre of the systems, velocity gradients along the filaments which may be interpreted as longitudinal gas flows that bring material to the centre of the system at rates of of $\sim10^{-4}$--$10^{-3}$\,\Msun\,yr$^{-1}$ \citep[e.g.][]{Peretto+14, Kirk+13, Williams+18, Lu+18, Chen+19, Anderson+21}. Whether the gas flows arise as a result of global hierarchical gravitational collapse, \citep[e.g.][]{Vazquez-Semadeni+19} or supernova feedback-powered inertia \citep[e.g][]{Padoan+20} is an ongoing matter of debate, but such large flow rates are clearly sufficient to transport a significant quantity of material into the centre of the region over a few Myr, and may be a crucial aspect of high-mass star formation.

HFSs have most commonly been identified within IRDCs in mid-IR survey data \citep{Peretto+Fuller09} which have superior resolution offered compared to sub-millimetre or millimetre survey data; for example, the 8\,\mic\ GLIMPSE survey data \citep{Benjamin+03} have a resolution of 2 arcseconds compared to 18 arcseconds of the APEX Telescope Large Area Survey of the Galaxy \citep[ATLASGAL;][]{Schuller+09}. IRDCs also make excellent targets for studies of the earliest phases of star formation due to their low level of emission at mid- and far-IR wavelengths. The relative level of 8\,\mic\ brightness has been shown to be a particularly effective tracer of the evolutionary state of molecular structures \citep[e.g.][]{Battersby+11, Rigby+21}, being dark at the early stages, and bright when star formation is advanced. This was quantified in a new metric, the infrared-bright fraction ($\firb$; \citealt{Rigby+21}; see also Watkins et al. in prep.), which measures the fraction of pixels within a structure that are brighter than the local background. \citet{Rigby+21} used this quantity to demonstrate that parsec-scale clumps within the Galactic Star Formation with NIKA2 \citep{Adam+18} Galactic Plane Survey (GASTON-GPS) tend to grow in mass at early times, supporting the gravitational collapse scenario for high-mass star formation as in \citet{Peretto+13}. 

The $\firb$ metric was used by \citet{Peretto+22} in conjunction with another new metric -- the filament convergence parameter, $\fc$, that quantifies the HFS-like nature of a clump by giving systems at the centre of radially-converging filaments a higher value -- to examine the evolution of HFSs in the same sample of GASTON clumps. The value of $\fc$ tends to increase with $\firb$, suggesting that either HFSs are a late stage of clump evolution, or that infrared-dark HFSs have short lifetimes compared to IR-bright HFSs, or both. A similar conclusion was reached by \citet{Kumar+20}, who found that high-mass star forming regions are systematically associated with hub-filament systems. The latter two studies are, crucially, based on single-dish data, and are limited to angular resolutions of $\gtrsim 10$ arcsec, which prevent core scales to be resolved for typical distances of a few kpc for high-mass star-forming regions.

At higher angular resolution, \citet{Anderson+21} compared the properties of 0.1-pc-scale cores within a sample of six infrared-dark HFSs to the cores within a larger sample of clumps mapped with ALMA, and found that infrared-dark clumps tended to have the highest efficiencies for formation of the most massive cores (MMCs), i.e. a greater fraction of the total clump mass was contained within the single most massive core, than for infrared-bright ones. Similarly, \citet{Traficante+23} found that the surface density of clumps and the MMCs increase with evolutionary stage. Studies from the ALMA-IMF large program \citep{Motte+22}, \citet{Nony+23} are finding a time dependence in the evolution of core mass functions (CMFs) in W43 \citep{Pouteau+23}, and \citet{Nony+23} found that the protostellar core mass function (CMF) in W43 is more top-heavy than the Salpeter-like pre-stellar CMF, indicating that the most massive cores accrete a larger fraction of mass than less massive ones. Taken together, these results suggest that the MMCs grow most rapidly at earlier times, while the rest of the core population catches up at later times as the clumps continue to gain material from their wider environment. However, we lack a detailed picture of how the dense gas in the centres of hub-filaments facilitate this growth of the most massive cores, and if and how these conditions differ from those found within non-HFS clumps, and across the range of masses. In this paper, we attempt to address this by examining the relationship between core masses and dense gas kinematics traced by \nthp\ across a sample of IR-dark clumps with a range of properties.

The fundamental rotational transition of the molecular ion diazenylium, \nthp, at 93 GHz was first detected by \citet{Turner74}, and later identified by \citet{Green+74}, and confirmed by \citet{Thaddeus+Turner75} and has since been widely utilised as a tracer of high column-density molecular gas. It is usually optically thin, and thus is ideal for studying the kinematics of dense gas linking them to the formation of dense cores in this study. Its rotational transitions -- including the $J$=1--0 transition at 93.174 GHz that we have observed in this study -- have at least seven hyperfine components \citep{Caselli+95}, though they often appear as a triplet in the interstellar medium (ISM) where linewidths are typically supersonic. The high-frequency $F_1, F = 0, 1$$\rightarrow$$1, 2$ component (sometimes called the `isolated component`) is located at a frequency equivalent to a doppler-shifted velocity of $-8$\,\kms, and a low-frequency triplet whose component centroids extend to $\sim+7$\,\kms\ with respect to a central triplet that defines the rest frequency. 

Due to this hyperfine structure, maps of the first and second moment of the full emission line can not be interpreted as simply as is the case for molecular emission lines that have no hyperfine splitting, such as CO. Performing fits of models to the hyperfine components in order to recover centroid velocities, linewidths and excitation temperatures has routinely been done within the literature with observations of \nthp; however, the high-spatial resolutions that are now achievable with facilities like NOEMA and ALMA have meant that multiple separate velocity components can be distinguishable within the spectra, which complicates fitting considerably. Some studies have achieved this by ignoring the majority of the hyperfine structure \citep[e.g.][]{Henshaw+14, Hacar+18, Barnes+21}, focusing their efforts on the isolated component, which may then be fit using a combination of a small number of Gaussian profiles. This is only achievable when the isolated components are clearly identifiable, which is often not the case in regions of very high column density and linewidth such as those within this study. Furthermore, the isolated component is a factor of a few weaker than the brightest component, except in optically thick regions, and so provides a reduced effective signal-to-noise ratio compared to fitting the full spectrum. To overcome these problems, we have developed a fully-automated multiple velocity component hyperfine line-fitting code to analyse \nthp\ cubes, called \mwydyn\ (the Welsh word for \textit{worm}, pronounced ``muy-din'', in recognition of the wiggly nature of the spectra).

In this paper, we examine the centres of a sample of IRDCs with existing NIKA \citep{Monfardini+10} and NIKA2 (GASTON-GPS) observations covering a range of properties, using high-resolution observations of \nthp\ (1--0) alongside several other key molecules and continuum, in order to examine the kinematics surrounding the most massive cores, and understand how the gas motions relate to the assembly of material within them. In Section \ref{sec:obs}, we described the sample selection and observations. In Section \ref{sec:analysis}, we describe the analysis of the data, including a description of a new fully-automated fitting code for lines with hyperfine structure, and extraction of core- and clump-scale properties. Our main results are described in Section \ref{sec:results}, and discussed in Section \ref{sec:discussion}, and we present our conclusions in Section \ref{sec:conclusions}.

\section{Targets and observations}
\label{sec:obs}

\subsection{Sample selection}
We compiled a sample of seven IRDCs from the catalogue of \citet{Peretto+Fuller09} to examine with NOEMA, which are a sub-sample of the study of \citet{Peretto+23}. The IRDCs were selected to cover a range of masses and morphologies, and to lie at comparable distances with $3.5 < d / \mathrm{kpc} < 5.0$ thus limiting the impact of a varying spatial resolution element. We also have overlapping 1.2\,mm observations from our NIKA \citep{Rigby+18} and NIKA2 \citep{Rigby+21} observing programs for these seven IRDCs. Systemic velocities for each of the IRDCs in our sample had been determined by observations of \nthp\ (1--0) in \citet{Peretto+23} using EMIR on the \emph{IRAM} 30-m telescope. The systemic velocities were then used to determine heliocentric distances using the batch input version (v2.4.1) of the \citet{Reid+16}\footnote{\url{http://bessel.vlbi-astrometry.org/node/378}} Bayesian Distance Calculator, which adopts the latest values of fundamental Galactic parameters from \citet{Reid+19}, including a Sun-Galactic Centre distance of 8.15 kpc. We make the following modifications to the configuration files: i) we insist that, since our targets are IRDCs, only the near kinematic distance solution is considered, by setting $P$(far) to zero; ii) since we do not know \textit{a priori} that the IRDCs reside within spiral arms, we disable the influence of the prior relating to the spiral arm model upon the distance probability density functions. The result is that our distances are primarily given by the near kinematic distance, but incorporating information for any maser sources with measured parallaxes. Details of the observations are given in Table~\ref{tab:targets}, including parsec-scale properties of the central clumps that are characterised in Section \ref{sec:clump_properties}.

\begin{table*}
\centering
\caption{Summary of observations, with columns as follows: 1) Source designations identify them in the catalogue of \citet{Peretto+Fuller09}; 2--3) coordinates of the NOEMA pointings; 4) systemic velocities from \citet{Peretto+23}; 5--6) heliocentric distance and uncertainty; 7--8) major and minor axes of the synthesised beam, and rms sensitivity of the 2.8\,mm continuum; 9--10) major and minor axes and rms sensitivity of the \nthp\ (1--0) observations; 11) total clump mass contained within a 1-pc-diameter aperture; 12) bolometric luminosity of compact sources within a 1-pc aperture; 13) infrared-bright fraction within the 1-pc aperture; 14) maximum filament convergence parameter within the 1-pc aperture. The top 7 rows are our new NOEMA observations, while the bottom 6 rows correspond to the adapted ALMA observations of \citet{Anderson+21}.}
\label{tab:targets}
\resizebox{\textwidth}{!}{
\begin{tabular}{cccccccccccccc}
\hline
Source & R. A. & Dec. & $v_\mathrm{sys}$ & $d$ & $\Delta d$ & $\theta_{2.8}$ & $\mathrm{rms}_{2.8}$ & $\theta_\mathrm{N_2H^+}$ & $\mathrm{rms}_\mathrm{N_2H^+}$ & $M_\mathrm{1pc}$ & $L_\mathrm{bol}$ & $\firb$ & $\fc$ \\
 & (J2000) & (J2000) & $\mathrm{km\,s^{-1}}$ & $\mathrm{kpc}$ & $\mathrm{kpc}$ & $\mathrm{{}^{\prime\prime}}$ & $\mathrm{\mu Jy\,beam^{-1}}$ & $\mathrm{{}^{\prime\prime}}$ & $\mathrm{mK}$ & M$_\odot$ & L$_\odot$ &  &  \\
 1 & 2 & 3 & 4 & 5 & 6 & 7 & 8 & 9 & 10 & 11 & 12 & 13 & 14 \\
\hline
SDC18.888$-$0.476 & 18:27:07.600 & $-$12:41:39.7 & 66.3 & 3.69 & 0.34 & $5.8\times2.6$ & 88.4 & $7.2\times3.2$ & 21.5 & 2950 & 2890 & 0.09 & 0.26 \\
SDC23.367$-$0.288 & 18:34:53.800 & $-$08:38:15.8 & 78.3 & 4.79 & 0.98 & $5.6\times2.8$ & 34.1 & $6.3\times3.3$ & 31.4 & 1740 & 1960 & 0.02 & 0.06 \\
SDC24.118$-$0.175 & 18:35:52.500 & $-$07:55:16.8 & 80.9 & 3.78 & 1.25 & $5.4\times2.9$ & 33.3 & $6.6\times3.4$ & 25.3 & 720 & 1020 & 0.11 & 0.23 \\
SDC24.433$-$0.231 & 18:36:41.000 & $-$07:39:20.0 & 58.4 & 4.97 & 1.38 & $5.5\times2.7$ & 65.5 & $6.8\times3.3$ & 29.4 & 3650 & 13260 & 0.47 & 0.53 \\
SDC24.489$-$0.689 & 18:38:25.600 & $-$07:49:37.3 & 48.1 & 3.85 & 0.85 & $5.3\times2.7$ & 99.1 & $6.3\times3.3$ & 41.6 & 890 & 350 & 0.00 & 0.39 \\
SDC24.618$-$0.323 & 18:37:22.800 & $-$07:31:38.1 & 43.4 & 3.82 & 0.96 & $5.6\times2.7$ & 40.1 & $6.9\times3.3$ & 33.3 & 650 & 4050 & 0.19 & 0.13 \\
SDC24.630+0.151 & 18:35:40.200 & $-$07:18:37.4 & 53.2 & 3.84 & 0.81 & $5.6\times2.7$ & 77.6 & $6.7\times3.2$ & 31.4 & 780 & 1340 & 0.07 & 0.18 \\
\hline
SDC326.476+0.706 & 15:43:16.331 & -54:07:13.5 & -38.0 & 2.14 & 0.43 & $5.5\times2.7$ & 151.0 & -- & -- & 2620 & 9430 & 0.36 & -- \\
SDC335.579-0.292 & 16:30:58.499 & -48:43:51.7 & -46.5 & 3.61 & 1.16 & $5.5\times3.6$ & 440.0 & --  & -- & 4360 & 50420 & 0.03 & -- \\
SDC338.315-0.413 & 16:42:28.133 & -46:46:49.3 & -39.4 & 2.44 & 0.52 & $5.5\times2.7$ & 55.1 & -- & -- & 540 & 1440 & 0.04 & -- \\
SDC339.608-0.113 & 16:45:59.439 & -45:38:40.3 & -32.9 & 2.21 & 0.37 & $5.5\times2.7$ & 120.2 & -- & -- & 1180 & 2000 & 0.19 & -- \\
SDC340.969-1.020 & 16:54:56.784 & -45:09:05.6 & -22.7 & 1.98 & 0.3 & $5.5\times2.7$ & 206.0 & -- & -- & 2810 & 3170 & 0.38 & -- \\
SDC345.258-0.028 & 17:05:12.373 & -41:10:02.5 & -16.4 & 1.33 & 0.12 & $5.5\times2.7$ & 67.1 & -- & -- & 330 & 20 & 0.26 & -- \\

\hline
\end{tabular}}
\end{table*}

\subsection{Observations}

The column density peaks of the seven IRDCs -- which we refer to herafter as the central \emph{clumps} -- were observed with single NOEMA pointings between 21st May and 30th August 2019 using the PolyFiX wideband correlator in Band 1 (3\,mm), providing $\sim$31\,GHz instantaneous bandwidth split over two sidebands each consisting of two polarizations. The local oscillator was tuned to a frequency of 99.5\,GHz, providing spectral windows covering approximately 88--96\,GHz and 103--111\,GHz. We selected several high-resolution spectral windows with 62.5\,kHz channel widths in order to provide 0.2\,\kms\ channels for our main target lines which were the (1--0) transitions of HCO$^+$, HNC, \nthp\ and C$^{18}$O.

A total of nine 15-m dishes were used in the 9D-configuration, which used stations W12, W09, W05, E10, E04, N13, N09, N05, and N02 with projected baselines ranging from 15 to 176\,m. The receiver bandpass was calibrated using the quasars 1749+096,1830-210, 3C454.3, 3C345, and 3C273. Phase and amplitude calibration was performed using observations of the quasars 1829-106, 1830-210, and 1908-201. The star MWC 349 was used as the primary flux calibrator, which is accurate to typically less than 10 per cent\footnote{\url{https://www.iram.fr/IRAMFR/GILDAS/doc/pdf/pdbi-cookbook.pdf}} in the 3\,mm band. Short-spacing observations for our spectral lines were obtained using the \emph{IRAM} 30-m telescope with dedicated observing runs in November 2019, and February, May, July, and October 2020. For the short-spacing observations we used EMIR with a setup optimised for our main target lines, with the local oscillator tuned to 99.5 GHz, and the FTS backend in 50 kHz mode to best match our NOEMA observations.

\begin{figure*}
    \centering
    \includegraphics[width=\textwidth]{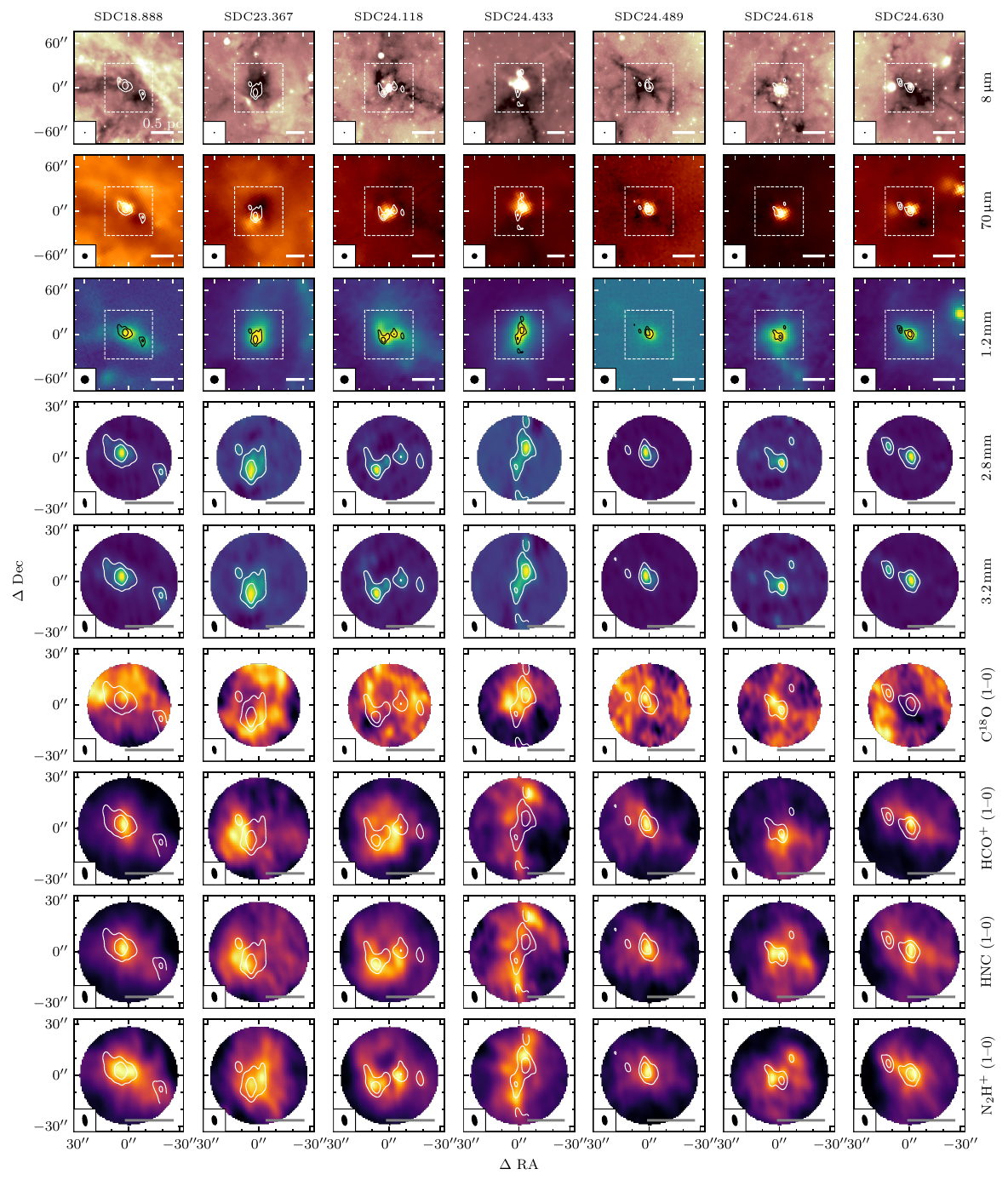}
    \vspace{-7mm}
    \caption{Each row shows all seven IRDCs in different bands. From top to bottom, the rows show: 1) \textit{Spitzer}/GLIMPSE 8-\mic\ continuum \citep{Churchwell+09}; 2) \textit{Herschel}/Hi-GAL 70-\mic\ continuum \citep{Molinari+16}; 3) 1.2\,mm continuum from NIKA for SDCs 18.888 and 24.489 \citep{Rigby+18}, otherwise from NIKA2 \citep{Rigby+21}; 4) NOEMA 107 GHz / 2.8\,mm continuum emission; 5) NOEMA 93 GHz / 3.2\,mm continuum emission; 6) NOEMA + \emph{IRAM} 30-m C$^{18}$O (1--0) integrated intensity; 7) NOEMA + \emph{IRAM} 30-m HCO$^+$ (1--0) integrated intensity; 8) NOEMA + \emph{IRAM} 30-m HNC (1--0) integrated intensity; 9) NOEMA + \emph{IRAM} 30-m N$_2$H$^+$ (1--0) integrated intensity. The beam size is shown in the lower-left corner of each image, and a scalebar representing a distance of 0.5\,pc is show in the lower-right corner. All coordinates are given with reference to the central coordinates given in Table \ref{tab:targets}. The top three rows (showing single-dish data) have a field width a factor of 2 larger than for the bottom six rows to provide more environmental context, with the smaller fields of view displayed as dashed boxes. Contours are shown for the 2.8\,mm continuum for each image at levels of 10 and 50 times the rms value.}
    \label{fig:targets}
\end{figure*}

The 2.8 and 3.2\,mm continuum images, and integrated intensity images from the main targeted lines are shown in Fig. \ref{fig:targets}, along with \textit{Spitzer}/GLIMPSE 8\,\mic\ \citep{Benjamin+03}, \textit{Herschel}/Hi-GAL 70\,\mic\ \citep{Molinari+16}, and \emph{IRAM} 30-m/NIKA and GASTON 1.2\,mm \citep{Rigby+18, Rigby+21} continuum images showing the wider environment.


\subsection{Data reduction}

Data reduction was performed using software from the {\sc gildas} suite\footnote{\url{http://www.iram.fr/IRAMFR/GILDAS}}. The raw data were calibrated using the {\sc clic} software, resulting in calibrated \textit{uv}-tables, which were then imaged and cleaned using {\sc mapping}. For the continuum sidebands at 2.8$\,$mm and 3.2$\,$mm, the procedure involved, first, producing a cleaned cube in order to identify and filter any spectral lines within the band that may result in over-estimated continuum flux densities, before imaging the cube with filtered channels, and cleaning.

Pseudo-visibilities for the single-dish data were combined with the NOEMA $uv$-tables for the four main target lines before imaging. To clean the images, we used robust weighting with a `robust' parameter of 0.5 to achieve a compromise between point-source sensitivity and resolution. We did not use any cleaning support in order to prevent the introduction of unwanted bias into the resulting emission maps. Two continuum images per source were also produced following the same procedure, with the exception that there are no short-spacing observations. The half power primary beamwidth, which defines the field of view, is 54.1 arcseconds for the \nthp\ (1--0) and 3.2\,mm continuum observations, for which the synthesised beam is typically $6.6 \times 3.2$ arcsec across our targets. For the 2.8\,mm continuum observations, the average beam size is $5.6 \times 2.7$\,arcsec, with a field of view of 47.0 arcsec. The individual beam sizes and rms values for each target are listed in Table \ref{tab:targets}.

\subsection{Supplementary high-resolution continuum data}

We also supplement our NOEMA continuum observations with 2.9~mm continuum observations of six IRDCs mapped previously with ALMA, as described in \citet{Anderson+21}: SDC326.476, SDC335.579, SDC338.315, SDC339.608, SDC340.969, and SDC345.258. These supplementary observations have a greater spatial extent, and have a higher angular resolution (with beams of typically $3 \times 2$ arcseconds) than the observations presented in this study, having fully-mapped the clouds, and so we first processed the ALMA observations to be more directly comparable to our single-pointing NOEMA data: the ALMA data were spatially smoothed to the average beam size for our NOEMA observations. The observations were then resampled onto a pixel grid of the same pixel size and extent (i.e. cropping to the NOEMA primary beamwidth) of our NOEMA maps, replicating the same field-of-view as our 2.8~mm observations, with the equivalent of the single-pointing centres being placed at the peak intensity of the Hi-GAL 500~$\upmu$m observation \citep{Molinari+16} of the same cloud. This latter choice was made to determine the pointing centres in a similar way than was done for our NOEMA sources, which were based on the peak intensities from our 1.15~mm NIKA and NIKA2 data. The characterisitics of the resulting processed data, including beam sizes, can be found in Table \ref{tab:targets}.

\subsection{Ancillary data} \label{sec:ancillarydata}

We make use of 8\,\mic\ continuum imaging data from the \textit{Spitzer} Galactic Legacy Infrared Midplane Survey Extraordinaire \citep[GLIMPSE;][]{Churchwell+09}, which have an angular resolution of 2 arcseconds. We also make use of 70, 160, and 250\,\micron\ imaging from the \textit{Herschel} infrared Galactic Plane Survey \citep[Hi-GAL][]{Molinari+16}, and column density maps generated from these images from \citet{Peretto+16}. The 70, 160, and 250\,\mic\ data have angular resolutions of $\sim$8, 12, and 18-arcsec, respectively, and the column density maps have the same resolution of the 250\,\mic\ images. Finally, we made use of 1.15\,mm  NIKA2 \citep{Adam+18} imaging from the GASTON-GPS \citep{Rigby+21}, as well as the NIKA maps of SDC18.888 and SDC24.489 from \citet{Rigby+18}. Both the NIKA \citep{Monfardini+10} and NIKA2 data have an angular resolution of 11 arcsec.

\section{Analysis}
\label{sec:analysis}

\subsection{Core population} \label{sec:sourceextraction}

The population of compact sources, within each field in the 2.8~mm continuum data was extracted using {\sc astrodendro}\footnote{\url{https://dendrograms.readthedocs.io/en/stable/index.html}}, a {\sc Python}-based implementation of the dendrogram  \citep{Rosolowsky+08} segmentation method. Sources were extracted using a minimum valid flux density level of 3 times the rms noise value, with a minimum significance for structures of 2 times the rms noise value, meaning that the minimum peak intensity of 5 times the rms noise value. For each map, the rms noise value was first determined using emission-free regions of the cleaned image. Finally, sources were required to have an area equal to at least half of the beam area. The latter choice was made as opposed to requiring a full beam area because, by construction, the dendrogram technique always underestimates compact source sizes as a result of clipping the wings of the source profile after convolution with the telescope beam, an effect that is particularly pronounced for point sources near the detection limit, or in crowded regions. To alleviate this problem, aperture corrections were applied to the core fluxes that are a function of the source area and peak signal-to-noise ratio, which are detailed in Appendix \ref{app:aperture_corrections}. We also find that the inclusion of this minimum source size assists with breaking down over-sized sources during the construction of the dendrogram, and recovering compact sources more accurately in crowded regions. The 2.8~mm data were favoured over the 3.2~mm data for source detection due to both their slightly higher angular resolution, and the greater intensity of dust continuum emission, at the cost of a slightly reduced field of view.

The source extraction procedure produces a mask for each of the sources it identifies, and we used these masks to calculate the properties of interest: integrated flux density, peak flux density, signal-to-noise ratio, and radius. We are particularly interested in the dendrogram `leaves', which are smallest structures within which no further substructure is discernible. Hereafter, we will refer to these objects as `cores', although we note that the beam size gives a spatial resolution of $0.1$\,pc at the mean IRDC distance, and so these cores will almost certainly contain further unresolved substructures within them.

Core masses were calculated using:

\begin{equation}
    M_\mathrm{core} = \frac{S_\nu d^2}{\kappa_\nu B_\nu(T_\mathrm{d})},
\end{equation}

\noindent where $S_\nu$ is the flux density integrated over the source area, $d$ is the distance to the source, $B_\nu(T_\mathrm{d})$ is the Planck function evaluated at frequency $\nu$ and dust temperature $T_\mathrm{d}$, and the dust opacity is given by: $\kappa_\nu = 0.1  \, \mathrm{cm}^2 \, \mathrm{g}^{-1} (\nu /\mathrm{THz})^{\beta}$. This value incorporates a gas-to-dust mass ratio of 100, and is the typical value adopted by e.g. \citet{Marsh+15}, \citet{Peretto+16}, and \citet{Anderson+21}. For consistency with the latter study in particular, we adopt a value for the dust emissivity spectral index, $\beta = 2$.

To determine the dust temperatures, we follow the same approach as \citet{Anderson+21}. Firstly, for all cores we adopt the flux density-weighted colour temperature from the maps of \citet{Peretto+16} in which the temperature is estimated from the \textit{Herschel}/Hi-GAL 160/250-\mic\ colour, again assuming $\beta = 2$. This technique is limited to the resolution of the 250-\mic\ data, which is 18 arcsec, and so we may overestimate the temperature of cold cores compared to the ambient dust temperature on larger scales. For cores containing any embedded star formation, we estimate the core temperature from the 70-\mic\ luminosity of any associated compact sources (i.e. protostars) within the catalogue of \citet{Molinari+16}. The bolometric luminosity of compact sources is known to be strongly correlated with 70-\mic\ luminosity \citep{Dunham+08, Ragan+12}, and so we use the relationship of \citet{Elia+17}:

\begin{equation}
    L_\mathrm{bol} = 2.56 \left(\frac{S_{70}}{\mathrm{Jy}}\right)\left(\frac{d}{\mathrm{kpc}}\right)^2 \mathrm{L}_\odot,
    \label{eq:Lbol}
\end{equation}

\noindent to calculate bolometric luminosities, where $S_{70}$ is the integrated flux density of the associated 70-\mic\ compact source. For optically thin dust heated by an internal star, the temperature at a particular radius can then be determined \citep{Terebey+93}:

\begin{equation} \label{eq:Tint}
    T_\mathrm{d}(r) = T_0 \left(\frac{r}{r_0}\right)^{-q} \left( \frac{L_\mathrm{bol}}{L_0}\right)^{q/2}
\end{equation}

\noindent where $q = 2 / (4 + \beta)$, $\beta=2$, and $T_0 = 25$\,K is the dust temperature at a reference radius $r_0 = 0.032$\,pc for a star of luminosity $L_0 = 520$\,\Lsun. For the radius of our cores, we adopt the `effective' radius of a circle with the source area (i.e. the sum of the area of all pixels in the leaf), $r = R_\mathrm{eff} = \sqrt{A / \pi}$, deconvolved by the NOEMA beam.

\begin{table*}
\centering
\caption{Calculated properties of the cores identified in Section \ref{sec:sourceextraction}: designations ordered by integrated flux density for each target; integrated flux density at 2.8\,mm and 2.9\,mm for the NOEMA- and ALMA-observed sources, respectively; the aperture correction that has been applied to the integrated flux density; mass, with uncertainties corresponding to the 16th--84th percentiles of the Monte Carlo-sampled uncertainty distributions; beam-deconvolved equivalent radius; core temperature.}
\label{tab:cores}
\begin{tabular}{cccccccc}
\hline
Designation & R. A. & Dec. & $S$ & $f_\mathrm{ap}$ & $M_\mathrm{core}$ & $R_\mathrm{eq}$ & $T_\mathrm{core}$ \\
 & (J2000) & (J2000)  & $\mathrm{mJy}$ &  & \Msun  & pc & $\mathrm{K}$ \\
\hline
SDC18.888-MM1 & 18:27:07.89 & -12:41:36.93 & 38.8 & 1.00 & $202^{209}_{106}$ & 0.109 & 33.2 \\
SDC18.888-MM2 & 18:27:06.34 & -12:41:48.14 & 10.3 & 1.00 & $124^{128}_{65}$ & 0.100 & 15.8 \\
SDC18.888-MM3 & 18:27:08.46 & -12:41:30.27 & 2.7 & 1.74 & $30^{31}_{15}$ & 0.017 & 16.9 \\
SDC18.888-MM4 & 18:27:08.87 & -12:41:35.99 & 1.3 & 1.95 & $14^{15}_{7}$ & 0.035 & 17.3 \\
SDC23.367-MM1 & 18:34:54.03 & -8:38:20.95 & 16.0 & 1.00 & $240^{279}_{139}$ & 0.208 & 20.5 \\
SDC23.367-MM2 & 18:34:54.59 & -8:38:10.79 & 0.8 & 1.11 & $16^{18}_{9}$ & 0.053 & 15.9 \\
SDC24.118-MM1 & 18:35:53.01 & -7:55:23.59 & 6.0 & 1.00 & $69^{97}_{47}$ & 0.069 & 17.1 \\
SDC24.118-MM2 & 18:35:52.03 & -7:55:15.34 & 3.3 & 1.00 & $39^{54}_{26}$ & 0.095 & 16.9 \\
SDC24.118-MM3 & 18:35:51.30 & -7:55:17.35 & 1.5 & 1.06 & $18^{25}_{12}$ & 0.089 & 16.6 \\
SDC24.118-MM4 & 18:35:52.61 & -7:55:18.97 & 1.4 & 1.76 & $5^{7}_{3}$ & 0.016 & 52.6 \\
SDC24.433-MM1 & 18:36:40.74 & -7:39:14.10 & 13.3 & 1.00 & $95^{121}_{60}$ & 0.094 & 43.3 \\
SDC24.433-MM2 & 18:36:41.07 & -7:39:24.31 & 4.5 & 1.15 & $96^{123}_{61}$ & 0.051 & 16.0 \\
SDC24.433-MM3 & 18:36:40.95 & -7:39:42.26 & 2.2 & 1.17 & $49^{63}_{31}$ & 0.084 & 15.5 \\
SDC24.433-MM4 & 18:36:40.72 & -7:38:58.45 & 2.0 & 1.16 & $41^{52}_{26}$ & 0.050 & 17.0 \\
SDC24.489-MM1 & 18:38:25.68 & -7:49:34.72 & 37.1 & 1.00 & $348^{414}_{206}$ & 0.137 & 21.1 \\
SDC24.489-MM2 & 18:38:26.43 & -7:49:32.80 & 2.1 & 1.10 & $27^{32}_{16}$ & 0.056 & 16.1 \\
SDC24.618-MM1 & 18:37:22.86 & -7:31:39.89 & 7.7 & 1.00 & $44^{54}_{27}$ & 0.137 & 32.5 \\
SDC24.618-MM2 & 18:37:22.31 & -7:31:28.20 & 0.6 & 1.14 & $7^{9}_{5}$ & 0.046 & 17.0 \\
SDC24.618-MM3 & 18:37:22.16 & -7:31:41.33 & 0.3 & 1.99 & $3^{4}_{2}$ & 0.026 & 19.1 \\
SDC24.630-MM1 & 18:35:40.17 & -7:18:36.82 & 22.9 & 1.00 & $161^{188}_{94}$ & 0.126 & 27.1 \\
SDC24.630-MM2 & 18:35:41.07 & -7:18:29.89 & 8.3 & 1.00 & $100^{117}_{59}$ & 0.093 & 16.8 \\
SDC326.476-MM1 & 15:43:16.61 & -54:07:14.37 & 338.5 & 1.00 & $600^{690}_{344}$ & 0.119 & 37.4 \\
SDC326.476-MM2 & 15:43:17.90 & -54:07:32.24 & 11.0 & 1.07 & $50^{58}_{29}$ & 0.029 & 16.0 \\
SDC326.476-MM3 & 15:43:16.93 & -54:06:59.20 & 3.7 & 2.21 & $11^{13}_{6}$ & 0.003 & 22.7 \\
SDC326.476-MM4 & 15:43:14.30 & -54:07:26.87 & 1.9 & 1.43 & $9^{10}_{5}$ & 0.017 & 15.9 \\
SDC335.579-MM1 & 16:30:58.75 & -48:43:54.58 & 285.3 & 1.00 & $1274^{1757}_{844}$ & 0.185 & 40.5 \\
SDC335.579-MM2 & 16:30:57.25 & -48:43:39.76 & 62.1 & 1.00 & $246^{337}_{163}$ & 0.121 & 45.3 \\
SDC335.579-MM3 & 16:30:57.08 & -48:43:48.17 & 4.1 & 2.06 & $42^{57}_{28}$ & 0.009 & 19.3 \\
SDC335.579-MM4 & 16:30:58.39 & -48:44:10.34 & 3.3 & 2.24 & $42^{59}_{28}$ & 0.024 & 15.5 \\
SDC338.315-MM1 & 16:42:27.52 & -46:46:54.27 & 6.2 & 1.11 & $12^{14}_{7}$ & 0.031 & 45.0 \\
SDC338.315-MM2 & 16:42:28.08 & -46:46:49.46 & 3.4 & 1.32 & $18^{21}_{10}$ & 0.022 & 17.5 \\
SDC338.315-MM3 & 16:42:29.12 & -46:46:38.52 & 1.0 & 1.22 & $6^{7}_{4}$ & 0.033 & 16.3 \\
SDC338.315-MM4 & 16:42:29.68 & -46:46:31.57 & 0.5 & 1.73 & $3^{3}_{2}$ & 0.013 & 16.1 \\
SDC339.608-MM1 & 16:45:58.82 & -45:38:46.92 & 11.8 & 1.87 & $15^{17}_{8}$ & 0.009 & 53.4 \\
SDC339.608-MM2 & 16:45:59.42 & -45:38:45.02 & 11.6 & 1.93 & $11^{12}_{6}$ & 0.008 & 71.4 \\
SDC339.608-MM3 & 16:45:59.48 & -45:38:52.13 & 8.2 & 1.18 & $41^{45}_{22}$ & 0.024 & 15.8 \\
SDC339.608-MM4 & 16:45:59.19 & -45:38:36.16 & 7.4 & 1.27 & $35^{39}_{19}$ & 0.021 & 16.5 \\
SDC339.608-MM5 & 16:46:00.49 & -45:38:32.87 & 4.2 & 1.18 & $22^{24}_{12}$ & 0.024 & 15.3 \\
SDC339.608-MM6 & 16:45:58.69 & -45:38:26.46 & 2.3 & 1.80 & $11^{12}_{6}$ & 0.033 & 16.3 \\
SDC340.969-MM1 & 16:54:57.29 & -45:09:04.73 & 123.3 & 1.00 & $148^{161}_{81}$ & 0.041 & 46.5 \\
SDC340.969-MM2 & 16:54:56.12 & -45:09:01.57 & 42.8 & 1.05 & $244^{267}_{133}$ & 0.028 & 11.6 \\
SDC340.969-MM3 & 16:54:58.39 & -45:09:09.15 & 7.4 & 1.26 & $25^{28}_{14}$ & 0.020 & 17.7 \\
SDC340.969-MM4 & 16:54:55.02 & -45:09:12.55 & 2.4 & 1.73 & $10^{11}_{5}$ & 0.011 & 15.3 \\
SDC345.258-MM1 & 17:05:12.16 & -41:10:06.37 & 8.2 & 1.19 & $16^{17}_{8}$ & 0.014 & 14.5 \\
SDC345.258-MM2 & 17:05:10.90 & -41:09:52.05 & 6.0 & 1.05 & $12^{13}_{6}$ & 0.044 & 14.1 \\
SDC345.258-MM3 & 17:05:12.10 & -41:10:11.05 & 5.8 & 1.47 & $11^{12}_{6}$ & 0.010 & 14.6 \\
SDC345.258-MM4 & 17:05:13.94 & -41:09:49.76 & 1.0 & 1.80 & $2^{2}_{1}$ & 0.015 & 15.2 \\
\hline
\end{tabular}
\end{table*}

In this way, we determined the masses of a total of 47 cores, with 21 cores and 26 cores from the NOEMA and ALMA data sets, respectively, and we present their properties in Table \ref{tab:cores}. The cores range in mass between  $\sim$2 and 1300\,\Msun, where the latter value belongs to SDC335.579 MM1, a well-known high-mass core \citep{Peretto+13}. With 10\% and 5\% uncertainties on the integrated flux densities for the NOEMA and ALMA data, respectively, a 0.5\,K uncertainty on colour temperatures \citep[as recommended by][]{Peretto+16}, and errors on the distance determinations that are typically $\sim$1\,kpc, the uncertainties are dominated by a factor of 2 uncertainty on $\kappa_0$ \citep{Ossenkopf+Henning94}. The core masses are therefore thought to be accurate to within a factor of $\sim$2.

\subsection{Clump-scale properties} \label{sec:clump_properties}
We calculated the masses of the central clumps within the thirteen IRDCs from the \citet{Peretto+16} column density maps\footnote{The column density maps for SDCs 326, 335, and 340 contained a small number of blank pixels near the column density peaks due to saturation in either the 160 or 250\,\mic\ \emph{Herschel} bands. These were filled by interpolation using the  {\sc interpolate\_replace\_nans} function from {\sc astropy.convolution} with a single-pixel standard deviation Gaussian kernel.} in order to provide context of the wider environment for our observations. The mass was measured within a 1\,pc-diameter aperture centred on the positions given in Table \ref{tab:targets} -- which is approximately the diameter of the NOEMA primary beam FWHM for our pointings -- since it is difficult to describe another definition that works well in the varying backgrounds across the sample. This is the approximate size-scale of what are typically called `clumps' \citep[e.g.][]{Ellsworth-Bowers+15a,Urquhart+18,Elia+21}, though we will hereafter refer to this particular measurement as the `1-pc clump mass' ($\mpc$), as a reminder of our fixed-aperture calculation. Because this technique adopts a fixed aperture size, the measurement is equivalent to the average surface density. 1-pc clump masses are calculated according to:

\begin{equation}
    \mpc = \mu_\mathrm{H_2} m_\mathrm{H} \int N_\mathrm{H_2}^\prime \mathrm{d}A
\end{equation}

\noindent where the surface area element $\mathrm{d}A = d^2 \mathrm{d}\Omega$, for the source distance $d$ and pixel solid angle d$\Omega$. In this case, the column densities used have been background-subtracted by first subtracting the minimum value of the column density within the aperture from each pixel: $N_{\mathrm{H}_2}^\prime = N_{\mathrm{H}_2} - N_{\mathrm{H}_2\mathrm{, bg}}$. We adopt a value of 2.8 for the mean molecular weight per hydrogen molecule, $\mu_\mathrm{H_2}$, which is the result of assuming mass fractions of 0.71 for hydrogen, 0.27 for helium, and 0.02 for metals. The recovered 1-pc clump masses range from 330--4360 \Msun, corresponding to mean surface densities of $\Sigma_\mathrm{pc} \sim $0.1--1.2\,g\,cm$^{-2}$ across the aperture. For context, this range of densities is encompassed by the density range of 0.05--1.0\,g\,cm$^{-2}$ of high-mass star forming clumps found in ATLASGAL \citep{Urquhart+14a}, indicating that all of our sources are capable of forming high-mass stars.

One measure of the dynamic status of a clump is through the virial parameter, which measures the ratio of gravitational to the kinetic energy. A clump is in virial equilibrium when the gravitational energy is equal to twice the total kinetic energy, $2 E_\mathrm{k} = E_\mathrm{G}$. Following the formulation of \citet{Bertoldi+McKee92}, we calculate the virial parameter:
\begin{equation} \label{eq:Mclump}
    \avir = \frac{5 \sigma^2 R}{\mathrm{G} M},
\end{equation}
\noindent where $\sigma$ is the total linewidth (including thermal and non-thermal contributions), $R$ is the radius, and $M$ is the mass. The choice of a fixed radius in our methodology would result, in cases where the parsec-mass is contained within an area that is smaller than the fixed aperture, in an overestimated virial parameter. Therefore, we calculated an adjusted radius and mass for the virial parameter determination only. To do this, we first determined the lowest closed contour in the \nthp\ (1--0) integrated intensity. We adopted the radius of a circle with the same area $A$ enclosed by the contour, $R_\mathrm{ctr} = \sqrt{A / \pi}$. We then calculated the background-subtracted column density within this same contour from the \emph{Herschel}-derived column density maps as in Eq. \ref{eq:Mclump}, and term this $M_\mathrm{ctr}$, which we use as our mass measurement (and note that this measurement more closely resembles the usual `clump mass'). Finally, the linewidths $\sigma_\mathrm{ctr}$ were then obtained by fitting the \nthp\ (1--0) spectrum, averaged over the region covered by the same contour as for the mass determination. We performed the fit using the {\sc gildas: class} model, which is described in more detail in Section \ref{sec:mwydyn}, though we fit only a single component here. We list the values derived for the virial parameter determination in Table \ref{tab:avir}. 

\begin{table}
\centering
\caption{Quantities derived for the determination of the virial parameters as described in Section \ref{sec:clump_properties}.}
\label{tab:avir}
\begin{tabular}{ccccc}
\hline
Source & $\sigma_\mathrm{ctr}$ & $R_\mathrm{ctr}$ & $M_\mathrm{ctr}$ & $\alpha_\mathrm{vir}$ \\
& \kms & pc & \Msun & \\
\hline
SDC18.888-0.476 & 1.66 & 0.30 & 1220 & 0.79 \\
SDC23.367-0.288 & 1.07 & 0.45 & 1960 & 0.31 \\
SDC24.118-0.175 & 1.15 & 0.36 & 370 & 1.51 \\
SDC24.433-0.231 & 1.61 & 0.32 & 1350 & 0.72 \\
SDC24.489-0.689 & 1.13 & 0.36 & 650 & 0.82 \\
SDC24.618-0.323 & 0.90 & 0.36 & 370 & 0.92 \\
SDC24.630+0.151 & 1.12 & 0.34 & 470 & 1.05 \\
\hline
\end{tabular}
\end{table}

A clump in virial equilibrium has $\avir = 1$, though we note that equipartition of kinetic and gravitational energy occurs at $\avir = 2$, and thus any clump with $\avir \leq 2$ is considered to be gravitationally bound. We note that this formulation of the virial parameter incorporates a factor of order unity which accounts for a non-uniform and non-spherical mass distribution, though the equation is derived under the assumption of a uniform density sphere. This means that for a source with a radial density profile of $\rho(r) \propto r^{-2}$, equipartition of kinetic and gravitational energy occurs at $\avir = 3.3$.  For the seven NOEMA clumps in our sample for which we have \nthp\ (1--0) linewidths, virial parameters are in the range 0.3--1.5, and so they are all considered to be gravitationally bound in keeping with the \citet{Peretto+23} measurements of the same sources.

The bolometric luminosity of each clump is calculated using Equation \ref{eq:Lbol}, and using the total integrated flux density of all 70-\micron\ compact sources from \citet{Molinari+16} that lie within the 1-pc aperture. Bolometric luminosities calculated in this way vary from $\sim$15--50000\,\Lsun\ across the sample, with all but two of the thirteen sources having $L_\mathrm{bol} > 1000$\,\Lsun\ -- the limit at which massive young stellar objects and \ion{H}{ii} regions are associated with ATLASGAL clumps \citep{Urquhart+14a}, and roughly corresponding to an embedded B3 ($\sim$6\,\Msun) or earlier type star.

We also supplement these properties with two new quantifications of each clump's evolutionary status. Firstly, for all thirteen sources we calculate the infrared-bright fraction $\firb$, which determines the fraction of pixels within the clump at 8\,\mic\ (from \textit{Spitzer}/GLIMPSE imaging; \citealt{Churchwell+09}) that are brighter than the local background within a 4.8-arcminute-wide box, and which has shown to be a good tracer of relative evolution (see \citealt{Rigby+21}, Watkins et al. in prep.). Sources evolve from $\firb = 0$ at the earliest stages of star formation to $\firb = 1$ at the latest stages, although we note that absolute time taken to evolve through this sequence is probably a function of mean density (via the free-fall time) and therefore mass. 

Secondly, we calculate the filament convergence parameter $\fc$ as presented by \citet{Peretto+22} for the seven NOEMA sources that have been observed by NIKA \citep{Rigby+18} or NIKA2 \citep{Rigby+21}. The convergence parameter quantifies the level of local filament convergence associated with each pixel by identifying the filaments within a field of a given radius and then quantifying how close they are to being radially aligned with that pixel. In this case, the field radius is set to 39 arcseconds, corresponding to 1\,pc at a distance of 5.2\,kpc, which is the median distance for clumps in the GASTON field. The filaments are identified using the second derivative method of e.g. \citet{Orkisz+19}, which determines a map of topological curvature, and identifies pixels which have eigenvalues below a threshold of $-3$ times the local standard deviation as being associated with filaments. These filaments are then reduced to single pixel-wide `skeletons' using {\sc scikit-learn}'s {\sc skeletonize} routine. \citet{Peretto+22} formulated the convergence parameter –– which is calculated for a pixel with coordinates $(x,y)$ -- as:
\begin{equation}
    \fc(x,y) = N_\mathrm{fil} \frac{\sum_{i=1}^{N_\mathrm{pix}}\cos(\Delta \theta)}{C_n},
\end{equation}
\noindent where $N_\mathrm{fil}$ and $N_\mathrm{pix}$ are the number of filaments, and the number of filament pixels within the search radius, respectively, $\Delta \theta$ is the angle between the radial direction from position $(x,y)$ to pixel $i$, and the filament direction at pixel $i$. $C_n$ is a normalisation constant which ensures that a pixel at the convergence point of six parsec-long radially distributed filaments has a value of $\fc = 1$.

For the five sources falling within the GASTON-GPS field, we take the maximum value within the NOEMA fields of view from the convergence parameter map of \citet{Peretto+22}, and for the two sources covered by NIKA, we calculate new convergence parameter maps in exactly the same way from the 1.2\,mm maps of \citet{Rigby+18}. This parameter quantifies the HFS-like nature of a clump by identifying nearby filaments, and providing a high value $\fc \rightarrow 1$ for sources which are located at the converging point of several nearby filaments, or a low value of $\fc =0$ for clumps that do not have any local filaments pointing towards themselves. We find that the NIKA and NIKA2 data are perfect for determining $\fc$ because they represent a combination of sampling the long-wavelength part of a dust spectral energy distribution (SED) that is most sensitive to dust column density, and relatively high-angular resolution ($\sim 11$ arcsec), and so we are unable to determine an equivalent value for the ALMA sample of IRDCs. In contrast to the NIKA and NIKA2 data, the closest match in angular resolution from \textit{Herschel}/Hi-GAL \citep{Molinari+16} would be the 160 or 250\,\mic\ data at 12 and 18 arcsec-resolution, respectively, but the continuum emission at these wavelengths is much more sensitive to local variations in temperature, and so are less suitable for tracing dust column density. Conversely, the 500\,\mic\ data are more suitable for tracing dust column density, but have relatively poor angular resolution, at 36 arcsec.

\subsection{Automated multi-component \nthp\ line-fitting: \mwydyn}
\label{sec:mwydyn}

Our \nthp\ (1--0) observations provide our primary means of assessing the kinematics of the dense gas within our clumps. As discussed earlier, it is necessary to fit the full hyperfine structure for these spectra in order to determine the intrinsic velocity dispersion and velocity centroids, and in many cases we can discern multiple discrete components within the spectra. We have therefore devised a fully-automated fitting algorithm -- called \mwydyn\ -- to decompose these spectra with up to 3 distinct velocity components that will enable us to carry out our kinematic analysis. In principle, \mwydyn\ is extensible to any molecule with hyperfine structure, and has been developed with testing on ALMA data (Anderson et al. in prep.) as well as the NOEMA data used in this study.

The procedure follows that employed by {\sc gildas: class}\footnote{\url{https://www.iram.fr/IRAMFR/GILDAS/doc/html/class-html/node8.html}}, which fits an individual \nthp (1--0) spectral component with a model containing four free parameters, but is in principle extensible to any molecule with a comparable hyperfine structure (e.g. HCN, NH$_3$). The method assumes that each component of the hyperfine multiplet shares the same excitation temperature and linewidth, and that the opacity varies as a function of frequency with a Gaussian profile. The total opacity is the sum of the opacity of the $N$ hyperfine component profiles, which can be expressed as:

\begin{equation}
    \tau(v) = \tau_\mathrm{total} \sum_{i=1}^{N} r_i \cdot \exp \left[ -4 \ln 2 \left(\frac{v - v_\mathrm{cen} - \delta v_i}{\Delta v}\right)^2 \right],
\end{equation}

\noindent where $\tau_\mathrm{total}$ is the sum of the opacities at the individual component line-centres, $r_i$ is the fractional intensity of the $i^\mathrm{th}$ component (the sum of which is normalised to unity), $\delta v_i$ is velocity offset of the $i^\mathrm{th}$ component relative to the velocity of the reference component at $\vcen$, and $\Delta v$ is the FWHM linewidth common to all components. The total line profile is then given by:

\begin{equation}
    T_\mathrm{mb}(v) = \frac{T_0}{\tau_\mathrm{total}}(1 - \mathrm{e}^{-\tau(v)}).
\end{equation}

\noindent Analytically, $T_0$ is proportional to $\tau_\mathrm{total}$, scaled by a factor that encapsulates the line amplitude.

\begin{SCfigure*}
    \centering
    \includegraphics[width=0.75\textwidth]{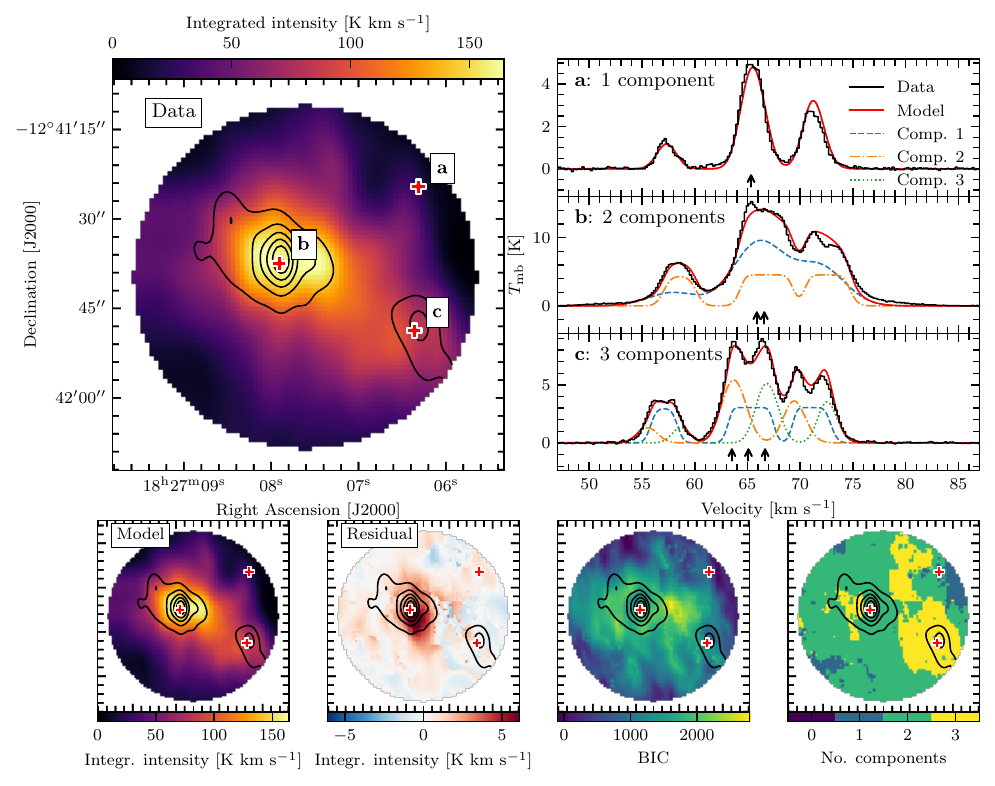}
    \caption{An illustration of the resulting fits from \mwydyn\ for \nthp\ (1--0) in SDC18.888 NOEMA. The top-left panel shows the integrated intensity, and is marked with 3 crosses corresponding to spectra shown in the top-right panel. The observed spectra are shown as steps, and the best-fitting model is shown as the solid line, along with its constituent components in dashed, dash-dotted, and dotted lines with centroid velocities indicated by arrows. On the bottom row, the left panel shows the model integrated intensity, with the residual image (data$-$model) shown in the middle-left panel. The middle right panel is a map of the BIC, and the right panel shows the number of velocity components for each spectrum. The black contours in all images show the 3.2\,mm continuum flux density, starting from 1 \mjybeam, with steps of 2 \mjybeam.}
    \label{fig:mwydyn_demo}
\end{SCfigure*}

Running \mwydyn\ on our \nthp\ (1--0) cubes results in the fitting of between one and three velocity components to each spectrum, with each fit being described by the four parameters: $\tau_\mathrm{total}$, $v_\mathrm{cen}$, $\Delta v$, and $T_0$. We detail the \mwydyn\ algorithm in Appendix \ref{app:mwydyn_procedure}, but in summary, the algorithm runs by:

\begin{enumerate}
    \item Determining an appropriate noise map.
    \item Cycling through each individual spectrum (i.e. pixel-by-pixel) whose peak intensity exceeds a signal-to-noise ratio threshold. During this process, initial guesses are first generated for a 1-component fit, which is then performed. Next, 2- and 3-component fits are attempted based on the result of the 1-component fit. Finally, the 1-, 2-, and 3- component models are compared, and the best fit is selected, provided that higher-component fits exceed a threshold of improvement over the simpler models.
    \item Comparing the fits to each spectrum with the fits of neighbouring spectra in order to determine if better solutions have been found locally.
    \item Write out data products.
\end{enumerate}

Figure \ref{fig:mwydyn_demo} illustrates some of the results from running \mwydyn\ on the \nthp\ (1--0) cube for SDC18.888, including a sample of spectra with their fitted components and combined model overlaid, and the output integrated intensity map alongside the residual image. We can see that \mwydyn\ produces a model that matches the data very well, with deviation in the integrated residuals on the order of $\sim$4\% at most. In the case of SDC18.888, emission is detected in every single pixel as a result of the pointing sampling only the centre of the IRDC, and so there is a model fit for every position in the image. The region of the largest integrated residuals is located just to the south-west of the brightest 3.2\,mm continuum emission source, indicating features in the spectral lines that were not perfectly modelled by \mwydyn, but it should be noted that this region is different to the location of the worst fits as quantified by the Bayesian Information Criterion (BIC) map. The component map also demonstrates that the largest number of components are not necessarily associated with the densest gas as traced by the continuum imaging.

\begin{SCfigure*}
    \centering
    \includegraphics[width=0.75\textwidth]{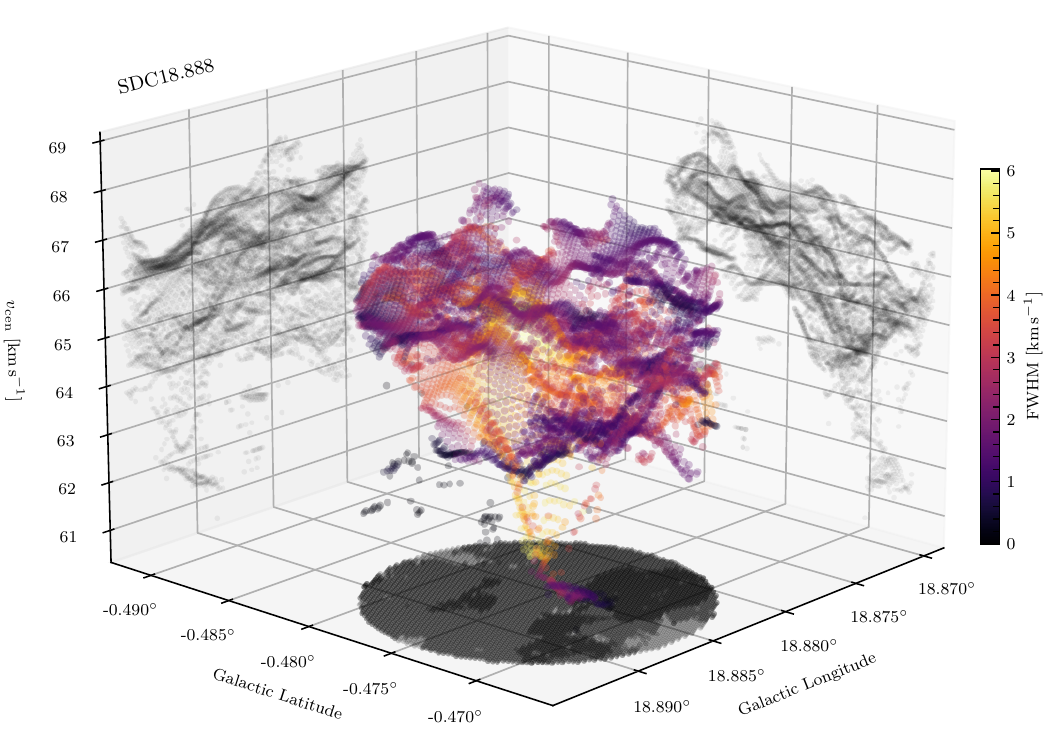}
    \caption{A three-dimensional representation of the fitting-results from \mwydyn\ for SDC18.888, in which the centroid velocity for each component is plotted against its Galactic coordinates. The points are coloured according to the FWHM linewidth of the component, with a transparency normalised by the integrated intensity of the line. Projections of the fitting results are also shown along each of the back surfaces as grayscale, in which the point colour is determined by the number of components along that axis. The lower surface shows the number of components fitted to each spectrum in grayscale, accordingly.}
    \label{fig:mwydyn3d}
\end{SCfigure*}

In Fig. \ref{fig:mwydyn3d}, we show a 3-D illustration of the fitting results to the full data cube for SDC18.888. In colour, the centroid of each component is plotted in ($\ell, b, \vcen$) coordinates, where the colour denotes the fitted FWHM of the component, and with a point opacity that is normalised by the integrated intensity of the component. Each surface illustrates a projection of the number of components along the three different axes. It is immediately obvious that the gas in the region is structured and highly complex. We note that high velocity dispersions between the multiple \nthp\ (1--0) emission components detected in each spectrum does not necessarily indicate the presence of structures at different physical separations, i.e. coherent structures identified in position-position-velocity (PPV) space do not always map onto coherent structures in 3-D space, and that the complexity in PPV space arises naturally in an inhomogeneous turbulent flow \citep{Clarke+18}. We display similar figures for the other six IRDCs in Appendix \ref{app:3dfigs}, and we interpret the results in Section \ref{sec:kinematics}.

\mwydyn\ is written in Python, and is fully parallelised. Run-times on these data were approximately 5 seconds per spectrum per CPU, on a computer cluster dating from 2016. The code is publicly available on GitHub\footnote{\url{https://github.com/mphanderson/mwydyn}}.

\subsection{Dense gas tracers} \label{sec:kinematics}
Our observational setup included a number of molecular species that probe different densities and conditions within our targets: \ceo, \hcop, HNC, and \nthp (1--0), and in this Section we present an overview of the general picture that they provide. In Fig. \ref{fig:targets}, we show the integrated intensity (moment 0) maps for these four molecular tracers alongside the continuum images. It is clear that \ceo\ (1--0) is tracing almost exclusively the material outside of the cores, which have a range of densities that barely overlaps a value of $10^4$\,\pcmmm\ where we expected CO freeze-out on to dust grains to be significant \citep{Bergin+Tafalla07}. The map exhibits little correspondence to the highest column densities traced by the 2.8 and 3.2\,mm continuum, though the cores are in fact visible in the \ceo\ (1--0) cubes, but the emission is relatively weak compared to the larger-scale emission. We consider this molecule to be a fairly accurate tracer of the clump envelopes. 

At the opposite end of the density scale, \nthp\ becomes detectable at moderate densities of $\gtrsim 10^4$\,\pcmmm\ \citep{Priestley+23}, and is generally optically thin. The \nthp\ (1--0) maps in Fig. \ref{fig:targets} are a much better match to the continuum images, but while there clearly is emission that is co-spatial with with continuum emission, it is also much more widespread. \nthp\ (1--0) is, therefore, tracing both the ambient clump material (at densities above the CO freeze-out) as well as the cores, and is valuable for tracing the transition from clump to core. It is important to recall here that the NOEMA data of all four of the molecular lines imaged here have been complemented with \emph{IRAM} 30-m observations to provide short-spacing information that allows the large-scale emission to be recovered. There are no such complementary observations for the continuum images, and so the differences in the morphology of the emission are caused by a combination of both chemical and observational (i.e. spatial filtering) effects.

\hcop\ and HNC show similar behaviour and are thought to trace similar conditions \citep[e.g.][]{Barnes+20, Tafalla+21}, and the maps are very similar, though the HNC maps also resemble a mix of both \hcop\ and \nthp\ (1--0) in several sources (e.g. SDC23.367 and SDC24.433). The critical density for both molecules is similar, with $n_\mathrm{crit} \sim 10^5$\,\pcmmm\ at 10\,K, but when accounting for more realistic excitation conditions and photon trapping, their effective excitation density is more like $\sim$$10^3$\,\pcmmm\ \citep{Shirley15}. They are not useful as tracers of column density as a result, however towards dense cores optically thick tracers like these can be useful as tracers of dynamic processes. For example, \hcop\ (1--0) often exhibits self-absorbed and blue-asymmetric spectra \citep[e.g.][]{Myers+96, DeVries+Myers05, Fuller+05}, which are a characteristic signpost of infall under gravitational collapse, and it is indeed useful in the sample for these purposes.

\begin{figure*}
\begin{subfigure}{0.31\textwidth}
    \centering
    \includegraphics[trim={0 0 0.8mm 0}, clip, width=\textwidth]{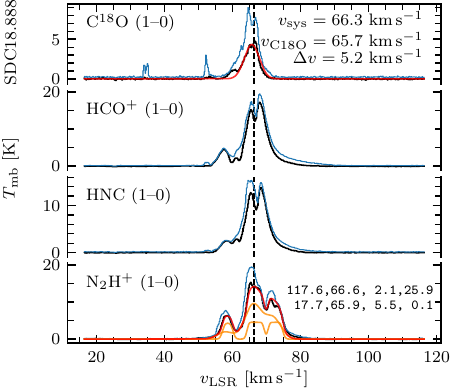} 
\end{subfigure}
\begin{subfigure}{0.31\textwidth}
    \centering
    \includegraphics[trim={0 0 0.8mm 0}, clip, width=\textwidth]{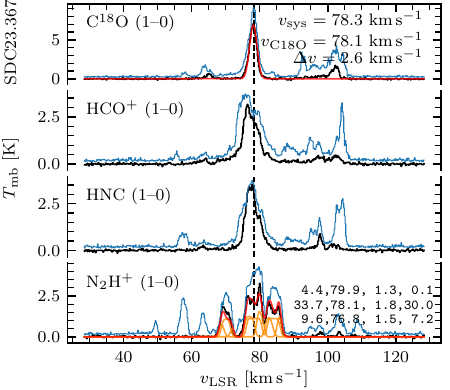} 
\end{subfigure}
\begin{subfigure}{0.31\textwidth}
    \centering
    \includegraphics[trim={0 0 0.8mm 0}, clip, width=\textwidth]{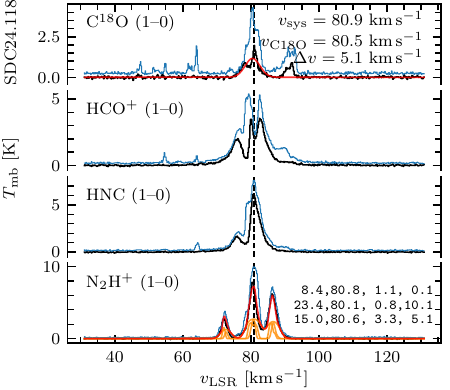} 
\end{subfigure}

\vspace{-5.3mm}
\begin{subfigure}{0.31\textwidth}
    \centering
    \includegraphics[trim={0 0 0.8mm 0}, clip, width=\textwidth]{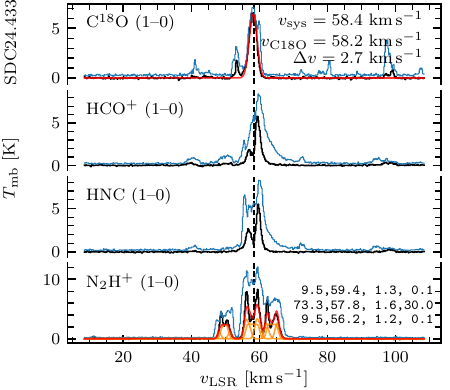} 
\end{subfigure}
\begin{subfigure}{0.31\textwidth}
    \centering
    \includegraphics[trim={0 0 0.8mm 0}, clip, width=\textwidth]{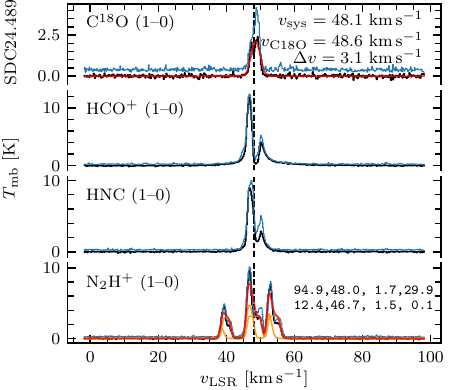} 
\end{subfigure}
\begin{subfigure}{0.31\textwidth}
    \centering
    \includegraphics[trim={0 0 0.8mm 0}, clip, width=\textwidth]{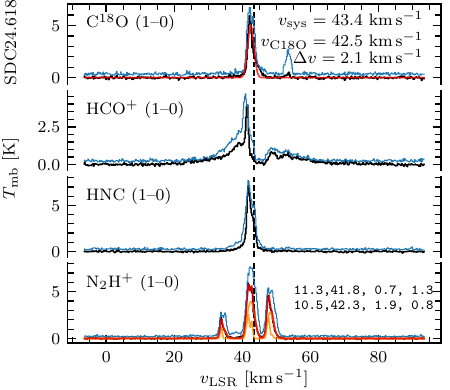} 
\end{subfigure}

\vspace{-5.3mm}
\begin{subfigure}{0.31\textwidth}
    \centering
    \includegraphics[trim={0 0 0.8mm 0}, clip, width=\textwidth]{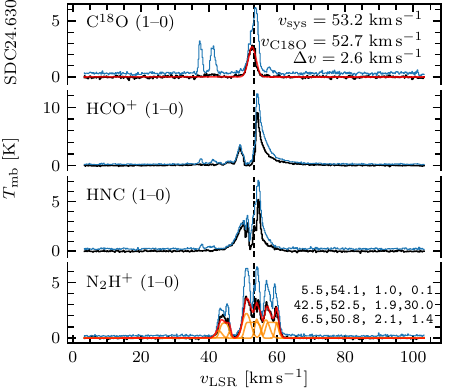} 
\end{subfigure}
\hspace*{-0.7mm}
\begin{minipage}{0.62\textwidth}
    \centering
    \hspace{7mm}
    \begin{minipage}{0.92\textwidth}
    \caption{Spectra of \ceo\ (1--0), \hcop\ (1--0), HNC (1--0) and \nthp\ (1--0) emission at the position of the brightest pixel of 2.8\,mm continuum emission for SDC18.888, SDC23.367, and SDC24.118 are shown in bold, while the thin spectra show the maximum value across all channels in the field. In the case of the \ceo\ spectra, a single Gaussian fit has been overlaid, and for the \nthp\ (1--0) spectra, we show the fits obtained with \mwydyn, where the parameter designation is given by the formalism in Section \ref{sec:mwydyn}, where {\texttt p1}, {\texttt p2}, {\texttt p3}, and {\texttt p4} refer to $T_0$, $v_\mathrm{cen}$, $\Delta v$, and $\tau_\mathrm{total}$, respectively. The systemic velocities identified by the single-dish data are shown as dashed vertical lines.}
    \label{fig:spectra}
    \vspace{50mm}
    \end{minipage}
\end{minipage}
\vspace{-42mm}
\end{figure*}

In Fig. \ref{fig:spectra} we show spectra for each of the targets. In all cases, we show the spectrum at the position of the peak of the 3.2\,mm continuum emission (the centre of the most massive core in all cases), as well as a secondary spectrum which shows the maximum pixel value across all channels in the field.  In the maximum spectra, all targets except SDC24.489 show velocities with emission features that do not correspond to the target clump, indicating that there are other molecular gas structures along the line of sight. This is unsurprising given the intermediate density range probed by \ceo\ (1--0), and the location of the targets in the inner Galaxy where the spiral structure of the Galaxy is relatively crowded in terms of velocity \citep[e.g.][]{Rigby+16}. We expect CO molecules to become depleted within the dense and cold molecular gas of cores, which may result in a drop in the optical depth that would allow the core systemic velocity to be distinguishable a peak. In general, the peak \ceo\ (1--0) emission at the position of the cores are well-modelled with Gaussian fits, but we do see such dips in the peak spectra for SDC18.888, SDC24.118, and SDC24.489, which may arise from self-absorption in optically thick gas, coupled with a temperature gradient along the line of sight. SDC24.118 is the clearest example of a complex doubly-peaked profile, while SDC24.618 shows the strongest degree of asymmetry.

The picture revealed by the \hcop\ and HNC (1--0) spectra is more complicated, though generally the two molecules share similar spectral features. We see blue-asymmetric spectra, possibly indicating infall motions, most clearly in SDC23.367, SDC24.489, and SDC24.618, while SDC18.888 shows blue asymmetric spectra in the most infrared-dark positions of the map, but not at the position of the continuum peak. In fact, the \hcop\ and HNC (1--0) spectra for SDC24.489 show the kind of self-absorbed and blue-asymmetric profiles that are archetypical infall spectra, and it is interesting to note that this is our most infrared-dark clump, with the highest $\fmmc$, and the second strongest hub-filament system as measured by $\fc$. Large wings in the spectra indicate the presence of outflows in most of these sources, and SDC24.489 is the outlier in this case too, with far less prominent outflow wings in the peak spectrum, though the red-shifted side of the spectrum does appear to show some outflow-like features. SDC24.618 is the only target in the sample in which the HNC (1--0) emission differs substantially from \hcop\ (1--0), and appears to be more closely related to the \nthp\ emission.

Our densest gas tracer, \nthp\ (1--0) has the most complicated spectra due to its hyperfine structure, and these show wide variation across the sample. The peak spectra in every case require at least two separate velocity components in the \mwydyn\ models indicating that the density structure is complex, even for SDC24.489 whose spectra are the most simple. SDC23.367 and SDC24.630 appear to show signs of self-absorption, which \mwydyn\ tends to model as three separate velocity components, with red-shifted and blue-shifted components either side of the central weaker component. The three sets of hyperfine lines tend towards a similar brightness temperature when the gas is optically thick, as is seen for these sources. In the case of SDC18.888, the \nthp\ (1--0) spectra also seem to be showing outflow wings that match the broad wings seen in \hcop\ and HNC (1--0).

\section{Results}
\label{sec:results}

\subsection{Core mass fractions}
\label{sec:cfe}

\begin{figure*}
    \centering
    \includegraphics[width=\textwidth]{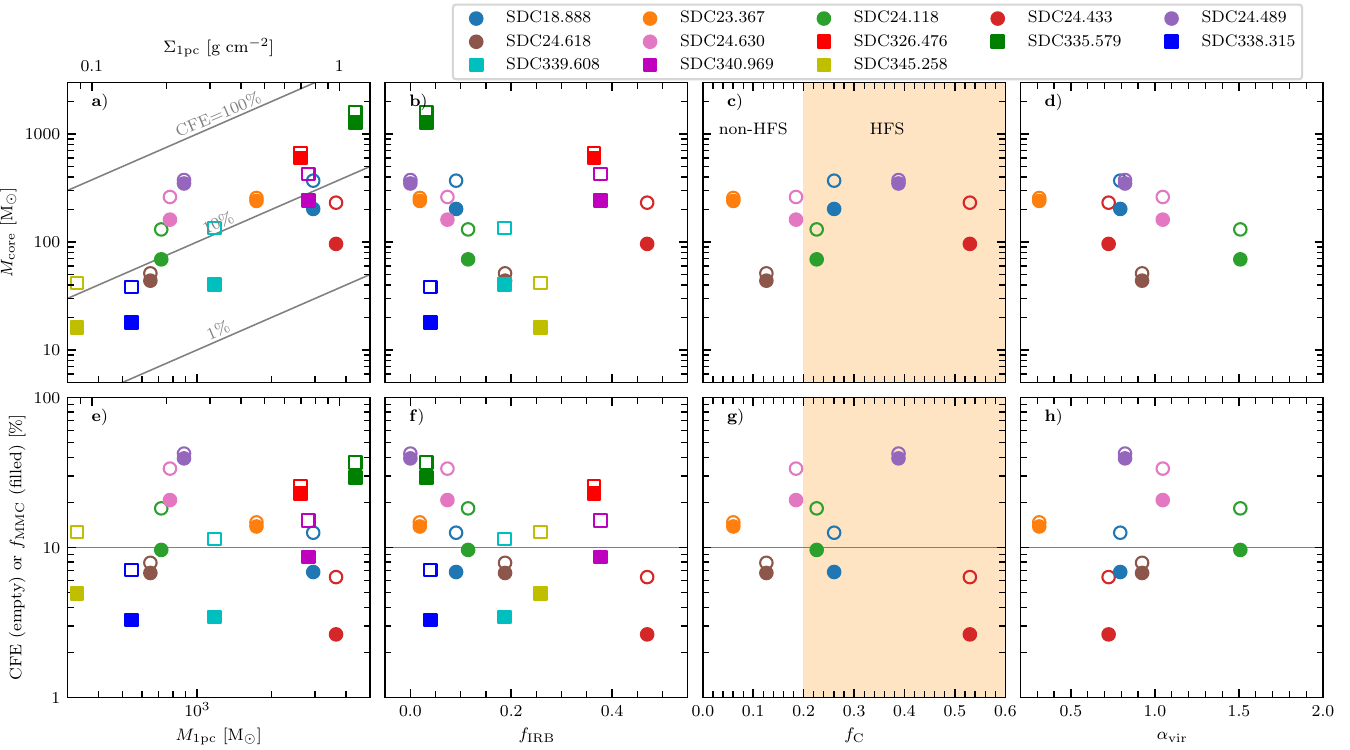}
    \vspace{-6mm}
    \caption{The top row shows 2.8\,mm core masses (solid symbols are MMCs, empty symbols are the sum of the mass in all cores) as functions of: a) 1\,pc mass (and its equivalent surface density on the secondary $x$-axis); b) the infrared-bright fraction; c) filament convergence parameter; d) virial parameter. The bottom row shows core formation efficiencies as functions of the same four parameters (solid indicating $\fmmc$, empty symbols for CFE). The colour coding shows the core populations for each IRDC, with IRDCs from the NOEMA and ALMA samples marked with circles and squares, respectively. The plots showing the filament convergence parameter have been shaded for values of $f_\mathrm{c} > 0.2$ which signifies the parameter space containing hub-filament systems \citep{Peretto+22}.}
    \label{fig:core_masses}
\end{figure*}

In this Section, we examine several ratios that characterise the relationships between the core populations and their clump-scale environments. The fraction of the 1-pc clump mass that resides within the most massive core is:
\begin{equation}
f_\mathrm{MMC} = \frac{M_\mathrm{MMC}}{\mpc},
\end{equation}
and similarly, the core formation efficiency (CFE) is the fraction of the 1-pc clump mass within its core population:
\begin{equation}
\mathrm{CFE} = \frac{\sum_i M_{\mathrm{core}, i}}{\mpc}.
\end{equation}
Finally, we utilise a metric that measures the relative dominance of the MMC over the rest of the core population:
\begin{equation}
q_\mathrm{MMC} = \frac{M_\mathrm{MMC}}{\sum_{i} M_{\mathrm{core}, i}}.
\end{equation}
We list these quantities, along with the number of cores per clump, in Tab. \ref{tab:MMcores}. In Fig. \ref{fig:core_masses}, we plot the core masses, $\fmmc$, and CFE -- as functions of 1-pc clump mass, $\firb$ -- indicating the evolutionary status of the clumps, and $\fc$ -- indicating the strength of HFS morphology. Both $\fmmc$ and CFE distributions are approximately log-normal, with a mean value of $\log_{10}(\fmmc) = -1.03 \approx 9\%$ and a standard deviation of 0.37 dex , while the CFE distribution has a mean value of $\log_{10}(CFE) = -0.80 \approx 15.7\%$ and with a standard deviation of 0.26 dex. We find a strong correlation between the 1-pc clump mass and MMC mass and total mass in cores, and we note that the quoted values for the correlations between the logarithms of these quantities are even stronger, indicating a power-law relationship. We detail the correlation coefficients for these pairs of variables, and for all of the relationships examined in this and subsequent sections in Fig. \ref{fig:correlations}, indicating those that are statistically significant.

\begin{table}
    \centering
    \caption{Core formation efficiencies for the MMCs ($\fmmc$) and for the sum of all cores (CFE), most-massive core-mass fractions ($\qmmc$), and the number of cores for each target in our sample.}
    \label{tab:MMcores}

    \begin{tabular}{ccccc}
        \hline
        MMC Designation & $\fmmc$ & CFE & $\qmmc$ & $n_\mathrm{cores}$ \\
        \hline
        SDC18.888-MM1 & 0.069 & 0.125 & 0.547 & 4 \\
        SDC23.367-MM1 & 0.138 & 0.147 & 0.939 & 2 \\
        SDC24.118-MM1 & 0.096 & 0.182 & 0.528 & 4 \\
        SDC24.433-MM2 & 0.026 & 0.063 & 0.415 & 3 \\
        SDC24.489-MM1 & 0.393 & 0.423 & 0.929 & 2 \\
        SDC24.618-MM1 & 0.068 & 0.079 & 0.855 & 2 \\
        SDC24.630-MM1 & 0.207 & 0.336 & 0.617 & 2 \\
        SDC326.476-MM1 & 0.229 & 0.255 & 0.896 & 4 \\
        SDC335.579-MM1 & 0.292 & 0.368 & 0.794 & 4 \\
        SDC338.315-MM2 & 0.033 & 0.071 & 0.466 & 4 \\
        SDC339.608-MM3 & 0.034 & 0.114 & 0.302 & 6 \\
        SDC340.969-MM2 & 0.087 & 0.152 & 0.570 & 4 \\
        SDC345.258-MM1 & 0.049 & 0.127 & 0.388 & 4 \\
        \hline
    \end{tabular}
\end{table}

In terms of core formation efficiency, we do not see any correlation between either $\fmmc$ or CFE and 1-pc clump mass. In Fig. \ref{fig:core_masses}e, we note a lack of $\fmmc$ values greater than 10 per cent for sources with -pc clump masses of less than $\sim700$\,\Msun, which might suggest that high CFE values are not obtainable if the local surface density is too low, though the number statistics here are very small. We see no correlation between the 1-pc clump mass, and the total number of cores per clump.

Examination of the CFE as a function of evolution, as traced by $\firb$, weakly suggests that the largest values are present at earliest times -- a trend that is seen both in the $\fmmc$ and CFE. The three sources that go against this trend --  SDC326.476, SDC338.315, and SDC340.969 -- are from the sample of IR-dark hub-filament systems of \citet{Anderson+21}. We find a weak correlation between $\firb$ and the number of cores per source, but the trend is not statistically significant given the sample size. Since $\firb$ traces only the relative evolution for clumps, and is expected to increase at different rates for clumps of different masses (with denser clumps having shorter free-fall times), it is possible that $L_\mathrm{bol}/\mpc$ may provide a more suitable tracer of evolution, and this quantity is used extensively in the literature for this purpose \citep[e.g.][]{Molinari+08, Urquhart+14a, Elia+17}. However, we see no correlation between $L_\mathrm{bol}/\mpc$ and any of the core mass-related properties, or the number of cores per clump reported in this Section. The $L_\mathrm{bol}/\mpc$ is indeed correlated with $T_\mathrm{core}$ within the sample, though the quantities are not always independent, since for cores that are associated with 70\,\mic\ compact sources, their temperatures are determined from their 70\,\mic-derived bolometric luminosities in Eq. \ref{eq:Tint}.

\begin{figure}
    \centering
    \includegraphics[width=\linewidth]{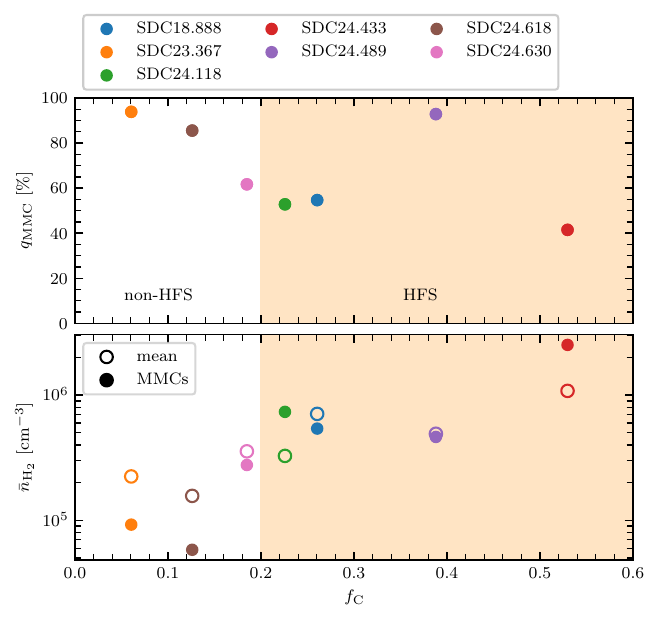}
    \caption{Top: The fraction of the total mass in cores that resides within the most massive core for the NOEMA clumps as a function of filament convergence parameter. Bottom: the mean density of cores (empty circles) and the mean density of the MMCs, per source, as a function of filament convergence parameter. Clumps with $\fc > 0.2$ are considered to be hub-filament systems.}
    \label{fig:MMCfrac}
\end{figure}

We see no link between either the filament convergence parameter $\fc$ or the virial parameter $\avir$ of the sources, and the associated masses or formation efficiencies of their constituent core populations. However, four out of the total of 13 IRDCs examined are dominated by a single core, which constitutes more than 75\% of the total mass in cores within the clump. We find that this quantity, $\qmmc$ shares a negative monotonic relationship with $\fc$, illustrated in Fig. \ref{fig:MMCfrac}, though this is not statistically significant unless the outlier SDC24.489 is excluded. With a standard dendrogram extraction, such a result could arise as a consequence of cores in a more crowded environment being separated at a higher contour level, artificially lowering their mass. However, we have mitigated this effect with aperture corrections that account for this splitting. We compare the mean densities of the cores to $\fc$, assuming the cores are spherical:
\begin{equation}
    \bar{n}_{\mathrm{H}_2} = \frac{M_\mathrm{core}} {\frac{4}{3} \pi R_\mathrm{eff}^3 \mu m_\mathrm{H}}
\end{equation}
\noindent where the mean molecular weight per hydrogen molecule $\mu = 2.8$. The densities range from $\sim 6 \times 10^4\,\mathrm{cm}^{-3}$ to $\sim4 \times 10^7\,\mathrm{cm}^{-3}$ and are strongly correlated with the convergence parameter, when considering the density of the MMCs, and all cores. This may be an indication of fragmentation into higher-density cores in the most strongly convergent systems.

We note that the various mass ratios used in this Section (CFE, $\fmmc$, most-massive core mass fraction) are distance-independent quantities, meaning that if the dust opacities also share a single value within clumps, that the uncertainties are very small. We present the quantities in Table \ref{tab:MMcores}.

\subsection{\nthp\ (1--0) kinematics revealed by \mwydyn} \label{sec:mwydyn_results}

In Sec. \ref{sec:mwydyn} we described our method for performing a multiple-velocity component fit to each spectrum of each \nthp\ (1--0) cube in our sample. In this Section, we discuss the results of this analysis, and how the gas kinematics relates to the positions of the cores seen in the continuum images.

\subsubsection{Interpretation of fitted parameters}
Before we examine the results of our \mwydyn\ fitting for the seven NOEMA IRDCs, it is important clarify some aspects of the resulting parameters. 

One of the key operating concepts of \mwydyn\ is that the emission within each \nthp\ (1--0) spectrum can be described as a linear (non-radiatively interacting) combination of 1--3 discrete parcels of gas with the various assumptions of the LTE-based fitting procedure (detailed in Section \ref{sec:mwydyn}), such as a single centroid velocity, velocity dispersion, and temperature for each component. In reality, the gas distribution of these regions is continuous across three-dimensional space, and with spatially varying densities over many orders of magnitude which are potentially radiatively coupled. Consequently, \mwydyn\ does not produce unique solutions for each model spectrum, but rather gives a combination of parameters which best fit the data, given the limitations of the model, and the parameter space that it is allowed to explore. We built in an approach where \mwydyn\ will prefer to model a spectrum in the simplest way, with the fewest components, and only add additional velocity components if the fit is significantly improved by them.

\mwydyn\ produces reliable results for spectra which have Gaussian component profiles that are not heavily blended, and this has been tested against synthetic spectra whose parameters can be accurately reproduced. However, real emission spectra are not always so well-behaved, and there are certain features that are present in our \nthp\ (1--0) cubes that are not captured by \mwydyn. Such complicating features include line-of-sight temperature fluctuations,  the presence of outflows, which reveal themselves by extremely broad wings in the profiles, and self-absorption, whose characteristic `m'-shaped spectral profiles \mwydyn\ tends to model as two or three distinct velocity components. In all of these cases, the spectral profiles will deviate from Gaussian distributions, and \mwydyn\ will compensate for these by adding extremely optically thick components, that are unlikely to realistically represent the gas properties, but that provide a better statistical fit. 

Due to the construction of our observations, which target the central parsec of the IRDCs in question, single-component fits tend to be more common in low-mass clumps, and towards the edges of the $\sim$1-pc fields (see Fig. \ref{fig:component_maps}). 2- and 3-component fits dominate the centres of almost all of our targets, which is an indication that the dense gas kinematics there are more complex, and these often consist of one or more extremely optically thick component (with $\tau_\mathrm{total} \lesssim 30$); we do not believe these to be accurate measurements of the optical depth of such components, and we caution against interpreting the values of $T_0$ and $\tau_\mathrm{total}$ in particular, too literally. Rather, the distribution of these parameters indicate that the gas is showing significant departures from single-temperate LTE conditions with Gaussian profiles. Since the same process is applied to all sources equally, we place the most emphasis on exploring the \emph{relative} differences between the fitted parameter distributions between, and within, our targets, and it is in this manner that \mwydyn\ is a powerful tool.

\subsubsection{Kinematic complexity}

\begin{figure}
    \centering
    \includegraphics[width=\linewidth]{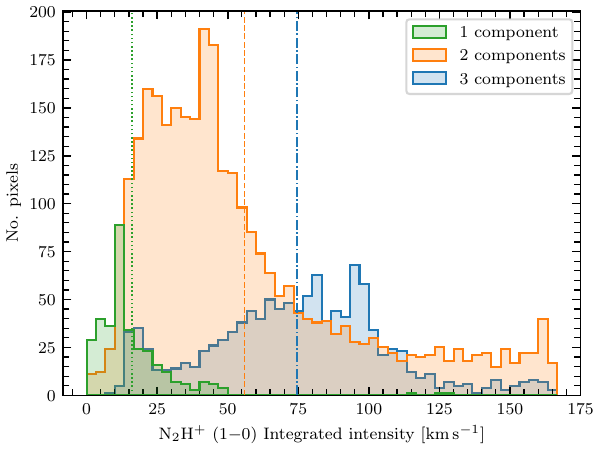}
    \caption{Distributions of \nthp\ (1--0) integrated intensity for pixels fitted with 1, 2, and 3-component \mwydyn\ fits for SDC18.888. The dotted, dashed, and dot-dashed vertical lines show the mean values for the 1-, 2-, and 3-component fitted spectra, respectively.}
    \label{fig:ncomp_integ}
\end{figure}

The \mwydyn\ fits reveal a generally high-level of what we will refer to here as `kinematic complexity', though `dynamical activity' might be an equally valid phrase. Our single-pointing observations towards these clump centres are not complete mappings of the IRDCs, as is evident in Fig. \ref{fig:targets}, but are probing the central 1\,pc, and consequently we are sampling what are probably the most extreme conditions within those clouds. In Fig. \ref{fig:ncomp_integ}, we show distributions of \nthp\ (1--0) integrated intensity per pixel for spectra that were fit with 1-, 2-, and 3-component \mwydyn\ fits. This illustrates, firstly, that the fraction of pixels in SDC18.888 that are fit by single components is small relative to the 2- and 3-component fits, and secondly that the higher-multiple component fits are found towards the spectra with the greatest integrated intensity. We also find that for the spectra with the greatest integrated intensities in Fig. \ref{fig:ncomp_integ}, these tend to be more often composed of 2-component fits as opposed to 3-component fits; this is where the blending of the spectral components has coupled with high-velocity wings that are so extreme that the 3-component fits do not managed to make a substantial improvement, and \mwydyn\ prefers the simpler models. The spectrum at position `b' in Fig. \ref{fig:mwydyn_demo} is a good example of this kind of spectrum.

We observe the same trend in all of the sources. In Fig. \ref{fig:kinematic_complexity}, we show that the mean number of \mwydyn-fitted components per spectrum ($N_\mathrm{comp}$) increases as a function of the 1-pc clump mass, and the 2-dimensional distributions of this can also be seen in row f) of Fig. \ref{fig:component_maps}. In general, we capture more of the quiescent outer edges of the sources with lower 1-pc clump masses, such as SDC24.118, SDC24.489, SDC24.618 and SDC24.630, for which a border of 1-component fits. For the most massive clumps, SDC18.888 and SDC24.433, we barely probe any quiescent regions, which we have essentially cropped out, resulting in a very small fraction of 1-component fits. Anderson et al. (in prep.) find this same trend of quiescent outskirts and complex interiors in a kinematic follow up to the six fully-mapped infrared-dark HFSs of \citet{Anderson+21}, with \mwydyn\ fits to the \nthp\ (1--0) data.

\subsubsection{Distributions of linewidths}
\begin{figure}
    \centering
    \includegraphics[width=0.95\linewidth]{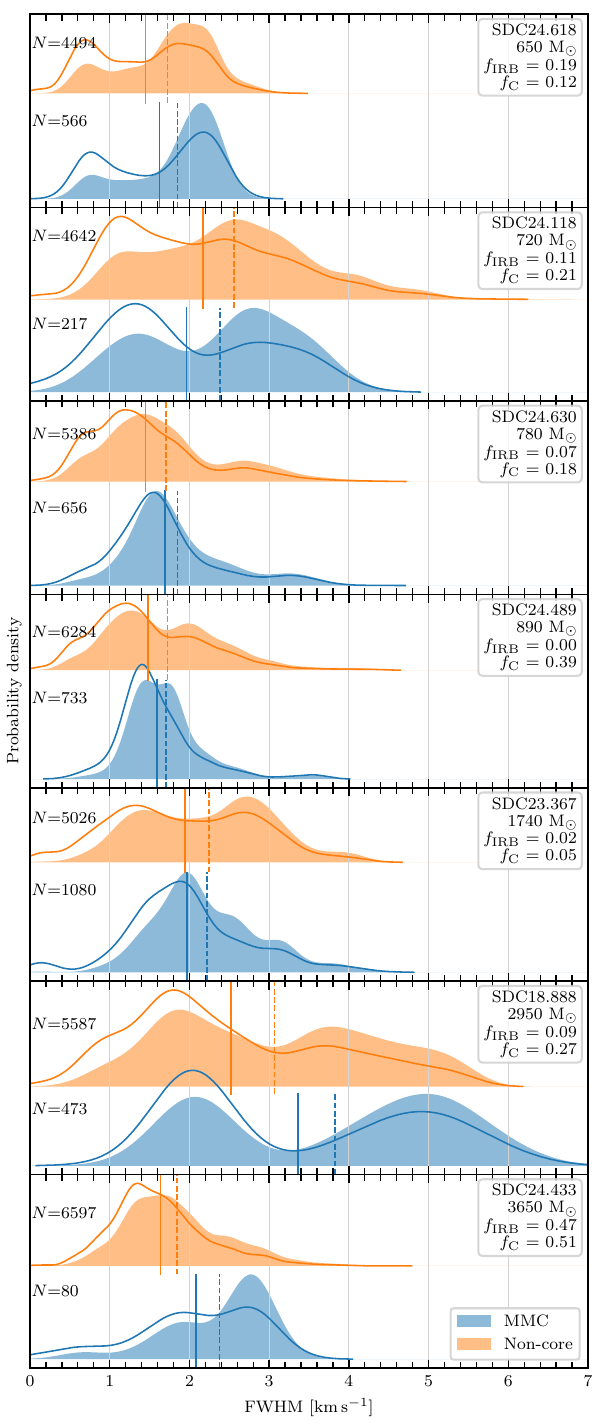}
    \caption{Gaussian kernel density estimators illustrating the distribution of the FWHM linewidths of components identified by \mwydyn\ for each source, ordered by 1-pc clump mass. Each data point in the filled distributions is weighted by the integrated intensity of the component, while the distributions shown in outline only were not weighted. The lower distributions for each source represent components within spectra that fall within the mask of the most massive core in each IRDC, while the upper distributions represent components that are not associated with any core. Solid and dashed vertical lines indicate the mean and weighted mean values, respectively. The number of components contributing to each KDE is displayed in the upper left.}
    \label{fig:FWHMKDE}
\end{figure}

To first order, and under the assumption that \nthp\ (1--0) is generally optically thin, the combination of the linewidths and the integrated intensities of the fitted components from \mwydyn\ encode the kinetic energy within the gas along a given line of sight. In Fig. \ref{fig:FWHMKDE} we compare the distribution of fitted linewidths from \mwydyn\ for pixels associated with the most massive core in each IRDC, and for pixels not associated with any core. We show these distributions using Gaussian kernel-density estimates (KDEs) to estimate the probability density functions. The shaded distributions have been weighted by integrated intensity (a proxy for mass), and we indicate the mean and weighted mean values for each of the distributions with solid and dashed vertical lines, respectively.

In several of the sources, SDC24.630, SDC24.118, and SDC24.618 the distributions that correspond to the MMCs have a greater fraction of spectra in the high-linewidth region of the distribution than the non-core distributions, as indicated by the relative positions of the (weighted and non-weighted) mean values. In some cases, such as SDC24.489 and SDC24.630, and 23.367, the MMC distributions are more sharply peaked at an intermediate linewidth, while the non-core distributions of linewidths are more widely spread, and we note that these are the three most infrared-dark clumps with the lowest values $\firb$. For SDC24.433, the MMC and non-core linewidth distributions are very similar, and in the case of SC23.367, the distributions are quite different, with a bimodal distribution for the non-core linewidths, and a skewed but singly-peaked distribution for the MMC. It is evident these distributions are systematically skewed to higher FWHMs in the weighted case than in the non-weighted case, indicating that higher column-density spectra tend to be more dynamic.

For each source, we test that the FWHM distributions for MMC and non-core components, and that the weighted and un-weighted components are significantly different using a two-sample Anderson-Darling test. The greatest similarity between any of the distributions is between the unweighted MMC and non-core samples of for SDC24.118, and even in this case, the $p$-value recovered is 0.005, and in all cases, $p \ll 0.05$, indicating that the null hypothesis that the two samples are drawn from the same underlying distribution can be rejected. However, the Anderson-Darling tests, with such large samples, become very stringent, and even very slight differences between distributions may result in very low $p$-values. 

There seems to be no single simple relationship that describes the behaviour in MMC-associated and non-core-associated pixels throughout the sample, and their interaction with clump-scale properties such as mass, evolution, and morphology. For example, the two most massive clumps, SDC18.888 and SDC24.433, have significantly higher weighted mean values in the MMCs than in the non-core distributions, but there is no clear mass-related trend amongst the rest of the sample. The position-position-velocity view of the clumps, seen in Figs. \ref{fig:mwydyn3d} and \ref{fig:3dfigs} show a high level of kinematic complexity. These 3D distributions exhibit many features, such as layers, gradients, sinusoidal oscillations within sub-structures, cavities, and loops. Such kinematic complexity is not captured by the distributions of linewidths, and so it is not surprising that the linewidth distributions alone are insufficient to elucidate the relationship between linewidths and clump-scale properties.

For four out of seven of these sources, the MMC distributions look similar in shape to the non-core distributions. SDC24.433, SDC24.489, and SDC23.367 have qualitatively differently-shaped distributions for the MMC spectra compared to the non-core spectra, but they too do not seem to share any particular set of characteristics, with a diversity of masses, $\firb$, and $\fc$ values (and this extends to the core-related statistics, $\fmmc$, CFE, and $\qmmc$). It is possible to come up with reasons for why these sources might be different; SDC24.433 is the most advanced in terms of evolution, with the highest $\firb$ and $L_\mathrm{bol}$ values, and so is more likely to contain strong stellar feedback that may results in higher linewidths. SDC24.489 is also an outlier in that it is a very strong hub-filament system, and very infrared-dark, with an anomalously high $\qmmc$ value (see Fig. \ref{fig:MMCfrac}). However, SDC23.367 does not appear to be remarkable in any particular way. Our sample is probably too small to identify the underlying trends that may be responsible for these similarities and differences. Despite these differences, we suggest that the distributions are not radically different, with similar mean and median values, and spanning a similar range, suggesting that the overall kinematics within the MMCs are not hugely different from the clump interiors, and that the MMCs are not kinematically decoupled from their parent clump at the scales probed by our observations.

\subsubsection{Global statistics} \label{sec:component_statistics}
To quantify the level of kinematic complexity, a number of further statistics, weighted by the integrated intensity, were calculated. As an initial step, we use an agglomerative clustering algorithm (from \texttt{sklearn.cluster}), to link together all components that are separated by less than 3 arcsec in the two spatial axes, and 0.5\,\kms\ in the spectral axis. We reduce our data to components that reside in the largest cluster identified in this way, in order to exclude any contamination from structures that are probably not spatially associated, but which fall along the same line of sight.

\begin{figure*}
    \centering
    \includegraphics[width=\textwidth]{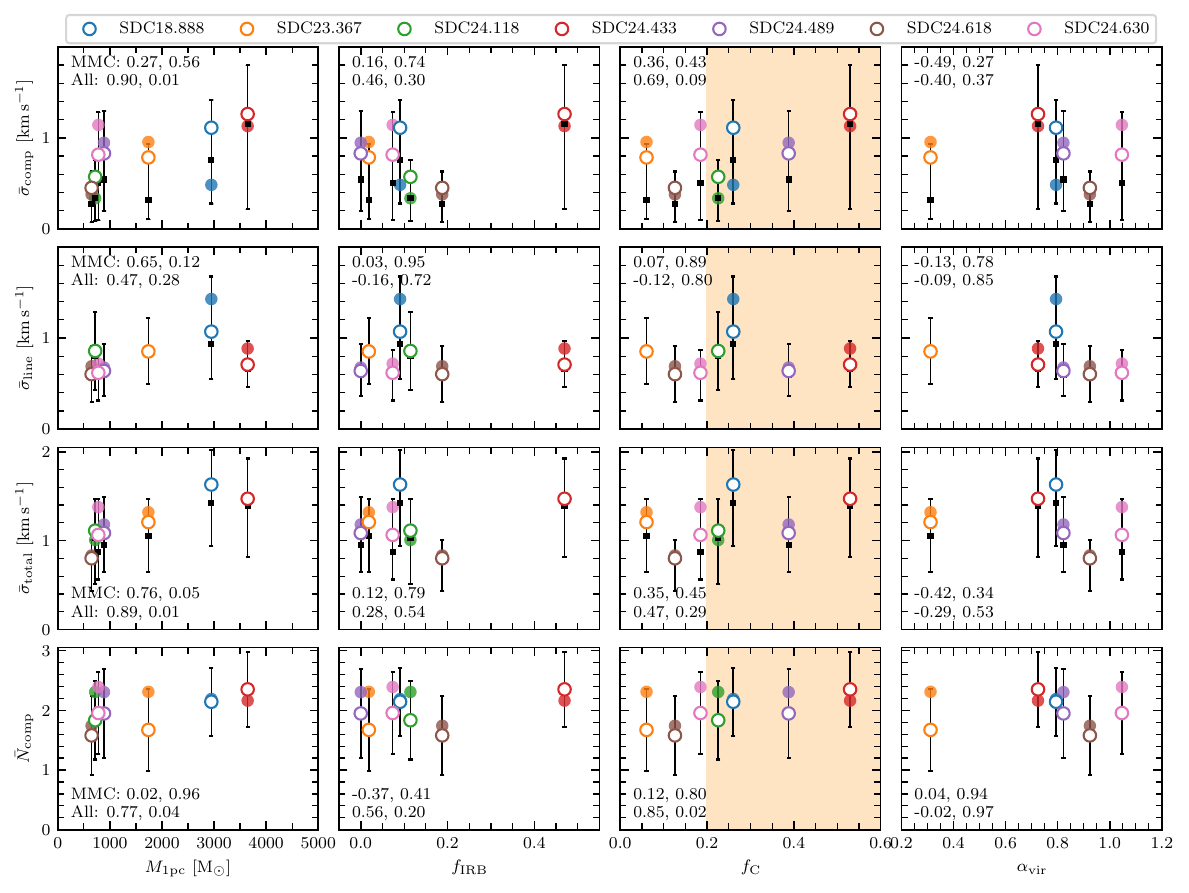}
    \caption{Summary statistics of dense gas kinematics (rows) as a function parsec-scale properties (columns) for each target. Top row: mean standard deviation of fitted centroid velocities; Second row: mean linewidth; Third row: mean total linewidth. Bottom row: mean number of fitted components per spectrum. These points are shown as functions of parsec-scale properties in each column, including: 1-pc mass, infrared-bright fraction $\firb$, filament convergence parameter $\fc$, and virial parameter $\avir$. In each case the empty circles indicate the values for the total distribution, while the solid circles show the values for the spectra associated with the most massive cores. For $\bar{\sigma}_\mathrm{comp}$, $\bar{\sigma}_\mathrm{line}$, and $\bar{\sigma}_\mathrm{total}$, the error bars show the range covered between the 16th and 84th percentiles of the full distribution, while the error bars for $\bar{N}_\mathrm{comp}$ give the standard deviation. Square markers, where visible, denote the median values. In all cases, the reported statistics (means, percentiles, and standard deviations) are weighted by the integrated intensity. For each panel, the Pearson correlation coefficients $\rho$ and $p$-values are given for the MMCs (top) and all spectra (bottom).}
    \label{fig:kinematic_complexity}
\end{figure*}

These statistics were calculated for $N_\mathrm{comp}$ total number of components within each target. With between 3918 and 4537 spectra per target, we note that the high sensitivity of our observations mean that we have detected and modelled at least one \nthp\ emission component in more than 85\% of spectra for every source. The total number of modelled components, $N_\mathrm{comp}$, ranges from 6341 to 10694 across the sample.

We first calculated the weighted mean centroid velocity over all components in each target:
\begin{equation}
    \bar{v} = \frac{\sum_i^{N_\mathrm{comp}} w_i v_i }{\sum_i^{N_\mathrm{comp}} w_i},
\end{equation}
\noindent where $w_i$ is the weight of the component, for which we adopted the associated integrated intensity. Similarly, we calculated the weighted mean linewidth:
\begin{equation}
    \bar{\sigma}_{\mathrm{line}} = \frac{\sum_i^{N_{\mathrm{comp}}} w_i \sigma_{\mathrm{line}, i}}{\sum_i^{N_{\mathrm{comp}}} w_i},
\end{equation}
\noindent where $\sigma_{\mathrm{line}, i} = \mathrm{FWHM}_i / \sqrt{8 \ln{2}}$), from our \mwydyn\ results. We next calculated the weighted dispersion between the individual velocity components with respect to the weighted mean centroid velocity:
\begin{equation}
   \bar{\sigma}_{\mathrm{comp}} = \sqrt{\frac{\sum_i^{N_{\mathrm{comp}}} w_i (v_i - \bar{v})^2}{\sum_i^{N_{\mathrm{comp}}}w_i}}.
\end{equation}
\noindent Finally, we calculated the weighted mean total velocity dispersion -- a quantity which encapsulates both the spread in centroid velocities about the global (weighted) mean, and the component linewidths:
\begin{equation}
\bar{\sigma}_\mathrm{total} = \sqrt{\frac{\sum_i^{N_{\mathrm{comp}}}w_i[(v_i - \bar{v})^2 + \sigma_\mathrm{line}^2]}{\sum_i^{N_\mathrm{comp}} w_i}},
\end{equation}

In Fig. \ref{fig:kinematic_complexity} we compare four quantities that summarise the level of `kinematic complexity' in each IRDC: global weighted mean values for the centroid dispersion, linewidths, total velocity dispersion, and number of components with the four quantities that describe the evolutionary state of the host clumps: 1-pc clump mass, $\firb$, $\fc$, and $\avir$. In each case we have calculated the Pearson correlation coefficients and $p$-values to test for linear correlations. We further calculated these same properties for only those spectra associated with the MMCs, for which we display the weighted mean values in Fig. \ref{fig:kinematic_complexity} too. Correlation coefficients are also presented in Fig. \ref{fig:correlations}. These weighted mean values are generally larger in the MMC spectra compared to the full sample, but this is not always the case.

Considering the statistics for all components (empty markers), we find that the general trend is for three of the four quantities, $\bar{\sigma}_\mathrm{comp}$, $\bar{\sigma}_\mathrm{total}$, and $\bar{N}_\mathrm{comp}$ to increase with 1-pc clump mass, while $\bar{\sigma}_\mathrm{comp}$ and $\ncomp$ are also correlated strongly with convergence parameter. None of the quantities are correlated with $\firb$. There are no correlations between the four \mwydyn-based component statistics and the virial parameters within our sample. When considering the difference between the MMC-only statistics and the global statistics, it is interesting that the most of the correlation between 1-pc scale and kinematic properties disappear, while the correlations between $\mpc$ and $\sigline$ and strengthens marginally. 

We have also performed the same calculations on a pixel-wise bases in order to produce 2-dimensional maps of these quantities, which we include in Appendix \ref{app:componentmaps}, and display in Fig. \ref{fig:component_maps}. We discuss our interpretation of these findings in Sec. \ref{sec:discussion}.

\subsection{Relationships between clump-scale, core-scale, and kinematic properties}

\begin{figure*}
\centering
    \includegraphics[]{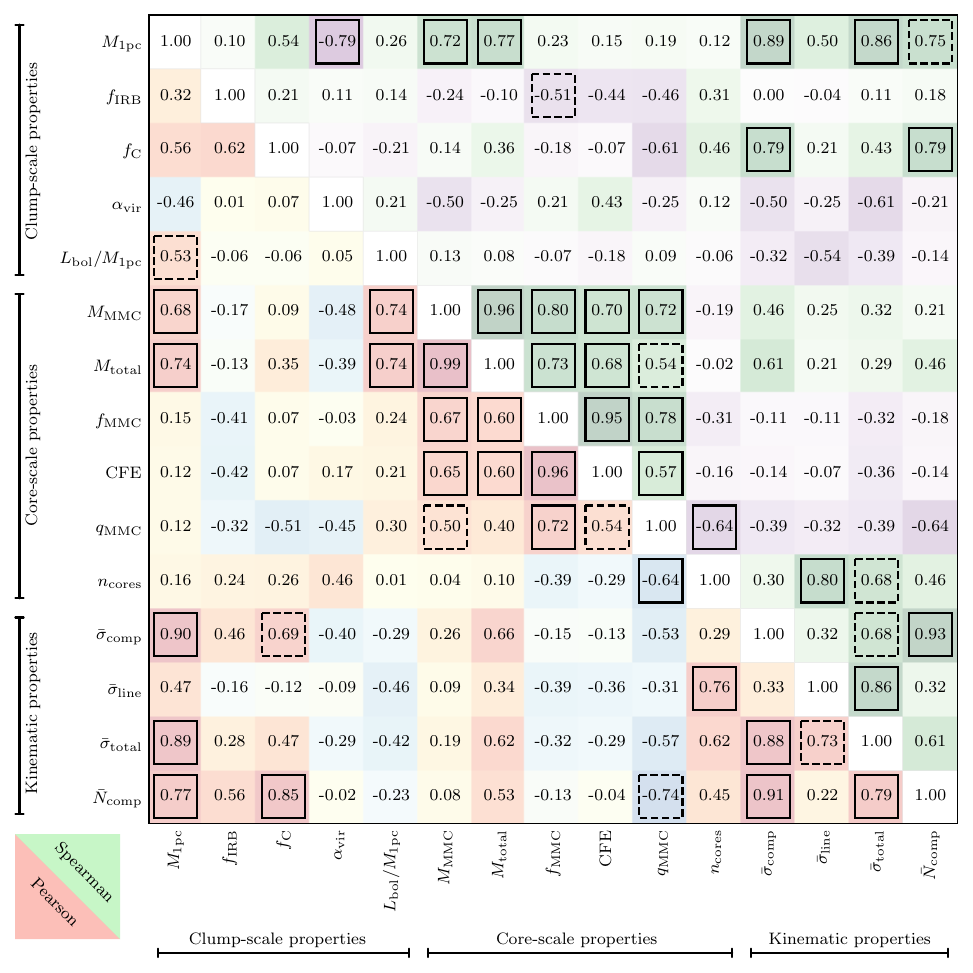}
    \caption{Pearson correlation coefficients ($\rho_\mathrm{P}$) and Spearman rank-order correlation coefficients ($\rho_\mathrm{S}$) for various combinations of quantities investigated in Section \ref{sec:results}. The $\rho_\mathrm{P}$ values are given to the lower-left of the diagonal, for which stronger red and blue colours indicate stronger positive and negative correlations, respectively, while $\rho_\mathrm{S}$ values are given to the upper-right of the diagonal, and stronger green and purple colours indicate stronger positive and negative correlations, respectively. The coefficients for correlations that are statistically significant (i.e. with $p$-values $\leq 0.05$) are highlighted with a solid box, while coefficients for correlations that have marginal significance ($0.05 < p \leq 0.10$) are highlighted with a dashed box.}
    \label{fig:correlations}
\end{figure*}

Thus far, we have examined the relationships between clump- and core-scale properties, and between clump-scale and kinematic properties. Here, we explore the relationships between all of the properties. Since it would be prohibitively tedious to examine individual figures concerning every pair of parameters, in Fig. \ref{fig:correlations} we present the Pearson and Spearman rank-order correlation coefficients ($\rho_\mathrm{P}$ and $\rho_\mathrm{S}$, respectively) between all pairings of the parameters presented thus far. The Pearson coefficients measure the strength of the \emph{linear} relationship between two variables, while the Spearman's rank measures the strength of \emph{monotonic} relationship that is not necessarily linear. We highlight those relationships which are statistically significant, given the $p$-values of the corresponding tests ($p \leq 0.05$), as well as those that are close to that limit ($0.05 \leq p \leq 0.10$) which may be of interest in future studies with larger samples. We note that some of these relationships have stronger and more statistically significant correlations in logarithmic space (resulting in larger $\rho_\mathrm{P}$ values) -- for example $\log_{10}(\mpc)$ vs. $\log_{10}(\mmmc)$ -- but we do not explore these here for the sake of simplicity.

The strongest relationships tend to be between properties of the same type; for example, core properties such as $\mmmc$ and $M_\mathrm{total}$ are most strongly correlated with the other core properties such as $\fmmc$ and CFE. With the exception of the clump-scale properties, this is neither surprising nor interesting because these variables are not independent. However, this illustration allows the identification of new relationships that may be of further interest. For example, sources greater weighted mean linewidths tend to be associated with a greater number of cores.


\section{Discussion}
\label{sec:discussion}

\subsection{The evolution of clumps}

\citet{Peretto+22} identified $\sim$2000 filaments within the $\sim2$\,deg$^2$ GASTON-GPS field, and examined the relationship between their orientation and the physical properties of the $\sim$1400 clumps within the field. They found that clumps with a higher value of the filament convergence parameter $\fc$ tend to be: i) more massive and ii) more infrared-bright (as measured by $\firb$) than clumps with a lower $\fc$ value. These suggest that HFSs are either a late-stage configuration in clump evolution, or that the evolution of HFSs is initially rapid, such that examples if IR-dark HFSs are relatively rare. The same survey data were also used to show that clumps accrete mass during the early stages of evolution \citep{Rigby+21}. These results provide an important context for the discussion of the results presented in this study.

On the 1-pc (i.e. `clump') scale, the properties of our sources broadly agree with this picture of clump evolution, with 1-pc clump mass being correlated with convergence parameter, and negatively correlated with virial parameter (albeit both correlations have low statistical significance, with Pearson's $p=0.19$ and 0.30, respectively). This is unsurprising given that five of the seven sources are within the GASTON-GPS field, but this would not be guaranteed to be the case given the dramatically smaller sample size. We do not see a correlation between clump mass and our tracer of evolution, $\firb$, but given the range in clump masses (and therefore free-fall timescales), $\firb$ is not expected to show a strong correlation.

In Section \ref{sec:cfe}, we found that both the total mass in cores, and the mass of the MMCs are strongly correlated with the 1-pc clump mass. \citet{Anderson+21} found a similar correlation between MMC mass and total mass in a sample of 35 clumps at various evolutionary stages, and found a much stronger correlation when limiting their sample to six IR-dark HFSs within the sample. \citet{Traficante+23} also report the same correlation in sample of 13 high-mass dense ($\Sigma > 1.0$\,g\,cm$^{-2}$) clumps at various stages of evolution. On the other hand \citet{Morii+23} did not find a correlation between MMC mass and clump mass in the ASHES sample of 39 70\,\mic-dark clumps, which are likely to be at an earlier stage in evolution than the aforementioned studies, though they do find that MMC mass correlates with clump surface density; due to our methodology, this latter result is consistent with the results presented here. When considering the various selection criteria in terms of evolutionary status of the targets, these results may all be consistent in a clump-fed scenario of star formation, where the cores' growth is promoted by the continuing accretion of material from the wider environment  \citep[e.g.][]{Peretto+20, Rigby+21}, and the correlation between MMC mass and 1-pc clump mass is weak at the earliest stages (i.e. 70-\mic-dark stages and 8-\mic-dark) and strengthens over time. Indeed, Fig. \ref{fig:core_masses} provides hints that the CFE is relatively high for the least evolved clumps in our sample with low values of $\firb$ (i.e. 70-\mic-bright but 8-\mic-dark), though the sample size is too small to really tell.

We also found that the fraction of the total mass in cores contained within the single most-massive core decreases as a function of $\fc$ (see Fig. \ref{fig:MMCfrac}). This quantity, $\qmmc$, may indicate the shape of the core mass function with a high value of $\qmmc$ indicating a top-heavy mass function within the clump centres, and vice versa (the quantity is clearly not, however, a robust measure of the complete pre-stellar core mass function within each source, as both the spatial extent and resolution are prohibitive in these observations). When coupled with the suggestion that $\fc$ increases over time for clumps, this indicates that mass accretion from the wider environment is initially concentrated on the MMC, with the surrounding cores receiving a greater fraction of infalling material at later stages. This does not necessarily suggest that the MMC stops accreting altogether, though this may well be the case since its early evolution is likely to be the most rapid, and therefore most likely to result in strong stellar feedback (indeed there are no 70-\mic-dark pre-stellar MMCs in our sample). Rather, the low $\qmmc$ values in the most hub-filamentary clumps simply suggest that infall or filamentary accretion rates towards the non-MMC cores are higher relative to the MMC at later stages, or this may indicate fragmentation of the MMC at later stages. SDC24.489 is an outlier in our sample for this, whose relatively high value of $\fc$ given its early evolution (with $\firb = 0$) appears to have resulted in an unusually high value of $\qmmc$ -- a point which we will return to.

The distributions of \nthp\ (1--0) centroid dispersions and, to a lesser extent, linewidths in Fig. \ref{fig:kinematic_complexity} also show a tendency to reach higher values for clumps which are more massive and have a higher $\fc$ value. These result in higher total dispersion values, though this trend is driven primarily by the inter-component dispersion per spectrum compared to linewidths. The general kinematic complexity of these regions increases with clump mass and $\fc$ too, indicated by the increasing mean values of the number of \mwydyn-fitted components per spectrum. These properties all negatively correlate with virial parameter, too, though the trends are not statistically significant. The relationship with virial parameter is complicated: if the most strongly bound clumps (with low virial parameters) are accompanied by infall signatures in a globally-collapsing cloud, the infall itself should increase the linewidth, counteracting the effect on the virial parameter \citep[e.g.][]{Larson81, Kauffmann+13}. The weaker correlations are, therefore, expected. The fact that clumps with greatest mass have the lowest virial parameters \emph{despite} their greater kinematic complexity indicates that gravity continues to dominate the overall balance of energy.

Overall, our measurements would seem to suggest that parsec-scale clump mass (or, equivalently, surface density) is the most important property in determining the dynamic properties of the dense gas in their interiors, and in determining the distribution of core masses. Hub-filament morphology, as measured by $\fc$, is a secondary factor, and probably one that is also driven by clump mass, with filaments converging more strongly towards more massive clumps over time \citep{Peretto+22}. We do not see any evidence for HFS having properties that separate them as a distinct sample from the non-HFS clumps in our sample, with no visible transition into different behaviour that occurs at $\fc = 0.2$. The most massive cores are located in the most massive clumps, where filament convergence is high, and with a high velocity dispersion between different velocity components of dense gas that requiring a higher level of complexity in \mwydyn\ models. This conclusion is in agreement with \citet{Kumar+20}, who suggested that all high-mass stars form in HFSs. 

\subsection{The dynamic centres of clumps}

The gas kinematics, as probed by our four main molecular tracers \ceo, \hcop, HNC, and \nthp\ (1--0) described in Section \ref{sec:kinematics}, consistently portray highly dynamic clump interiors. Despite our sample being chosen to cover the earliest IR-dark phases of clump evolution, and an absence in many cases of embedded objects at 8\,\mic, they all contain compact 70\,\mic\ sources, and display evidence for outflows, primarily visible in the \hcop\ (1--0) spectra, but also often in HNC (1--0) and even, in some cases, \nthp\ (1--0). Infall signatures are present in at least three of the seven NOEMA sources, and these tend to be concentrated towards the most 8\,\mic-dark regions, and are clearest in SDC24.489, our most infrared-dark cloud and second strongest hub-filament system (with $\firb$ = 0 and $\fc$ = 0.39). However, an absence of optically thick, self absorbed, and blue-asymmetric spectra does not guarantee the absence of infall. The classical \citet{Myers+96} infall profile assumes spherical symmetry in the density and temperature gradients, while infall occurring under more realistic and chaotic ISM conditions may result in a wide variety of spectral profiles, including red-shifted ones, depending viewing angle \citep{Smith+12a}, so it is quite possible that there is ongoing infall in the other clumps too. This result is broadly in keeping with \citet{Jackson+19}, who found that blue-asymmetric profiles were present in the majority of \hcop\ (1--0) profiles within a sample of $\sim1000$ high-mass clumps from the MALT90 survey, and that such profiles were more common at early stages of clump evolution.

The IRDCs in our sample were selected from the sample of  \citet{Peretto+23}, whose analysis of the velocity dispersion radial profiles on scales of a few tens of parsecs down to a few tenths of a parsec by showed that clumps located within these IRDC centres are dynamically decoupled from their parent molecular clouds, i.e. the radially averaged velocity dispersion as traced by \nthp (1--0) remains constant within the central few parsecs (i.e. the region probed by our observations). The proposed interpretation is that clumps are globally collapsing, while the rest of the molecular gas is not. In that context, the complex kinematics that is being seen in our \nthp\ (1--0) data in particular (as well as the other lines) has to be the result of intertwined gravity-driven mass inflows whose density structures vary according to the local conditions.

The dynamic conditions of the dense gas are most clearly visible in the results of our \mwydyn\ fits to the \nthp\ (1--0) spectra presented in Section \ref{sec:mwydyn_results}. In Fig. \ref{fig:FWHMKDE} we compared the distributions of linewidths in the MMCs and in non-core spectra, and they are qualitatively fairly similar, indicating that the MMC kinematics are not radically different from the surrounding gas. The points $\sigtotal$ points in Fig. \ref{fig:kinematic_complexity} representing the condensed \mwydyn-based total linewidths for the MMC spectra are almost universally located at only marginally higher values than the overall distributions, indicating that the level of kinematic complexity is somewhat elevated in the most massive cores. Overall, we suggest that this represents a level of kinematic similarity between the ambient clump material and the gas material, that may be expected if the cores are continually accreting material from the wider environment, as suggested by many other studies \citep[e.g.][]{Peretto+20, Rigby+21}, with both environments being highly complex, with supersonic linewidths. We temper this conclusion with the caveat that \nthp\ (1--0) may not be accurately tracing the kinematics of the most dense gas associated with the cores. Observations of higher density-tracing species such as N$_2$D$^+$ (1--0) may reveal a more accurate picture of the kinematics at core scales. Conversely, the kinematics of the diffuse gas in the clump and core environments (associated with outflows, for example) will not be represented in these distributions.

As mentioned in Section \ref{sec:mwydyn_results}, the structures revealed in position-position-velocity in Figs. \ref{fig:mwydyn3d} and \ref{fig:3dfigs} are incredibly diverse. One of the most striking features is the pervasive apparently sinusoidal oscillations that are common across our sample, and which are similar to structures that have been identified elsewhere in the literature. \citet{Hacar+13} identified similar structures (`fibres') within \ceo\ (1--0) observations at similar spatial scales to this study in the nearby filamentary complex L1495/B213 \citep[see also][]{ Tafalla+Hacar15}, and in \nthp\ (1--0) in the Orion Integral Shaped Filament \citep{Hacar+18}, and \citet{Henshaw+20} have identified such features across a wide range of scales, from 0.1-pc to kpc within the Milky Way and nearby galaxies. The possible explanation for these features is varied, with \citet{Henshaw+14} finding that they can be reasonably modelled as the result of infall or outflow.  These kinds of structures are clearly visible in our data within Figs. \ref{fig:mwydyn3d} and \ref{fig:3dfigs} in the grayscale two-dimensional position-velocity projections on the back of each axis, as well as within the coloured data points in the centre. Manipulation of the three-dimensional data suggests that, at least in our IRDC-centre observations, these structures may arise through the projection of concave and convex undulations within sheet-like layer; the oscillatory behaviour visible in 2-dimensional projects appear to be representing undulating velocity motions in 3-dimensional space that do not have a preferred spatial axis, and it is intriguing to note the lack of apparent filamentary structures (or fibres) on these scales in 3-D. The \nthp\ (1--0) fibres found in Orion \citep{Hacar+18} have lengths of $\sim$0.15\,pc and widths of $\sim0.03$\,pc, and so if similar structures were present in our targets, we might reasonably expect to detect (but not resolve) their presence. The close packing of the fibres may cause a level of confusion in data at lower physical resolution, such as ours, so the sheet-like structures in our observations might be a manifestation of this.

Four of the seven NOEMA sources surpass the threshold for being classified as hub-filament systems with $\fc > 0.2$ \citep{Peretto+22}, and so it is not clear whether we should expect to observe filaments within the central 0.5-pc of such systems. SDC24.618 is our lowest-mass clump, with the second-lowest filament convergence parameter ($\fc = 0.12$) and this is our best candidate for a filamentary or fibrous system. While the general structure in this clump also follows the sheet-like interpretation of the apparent fibres that the eye is drawn to in Fig. \ref{fig:3dfigs}, where they arise mainly as the projection of convex and concave structures in PPV space, there are a small number of isolated and velocity-coherent structures with large aspect ratios that could be considered to be filaments. If it is generally true that more massive clumps tend to evolve towards a hub-filamentary configuration, then in the very centres of these systems (which our observations target) we might expect the filaments to merge, eradicating the evidence for pre-existing fibres that make those filaments.

\subsection{The formation of cores in clump centres}

The values of $\fmmc$ and CFE are consistent with being drawn from log-normal distributions centred on 9 and 16 percent, respectively. The presence of these log-normal distributions indicate that the individual values may be determined by a series of random processes, and we attribute this to the chaotic clump interiors discussed in the previous Section, which are generated by conversion of gravitational potential energy to kinetic energy during the global collapse of the clump. In this interpretation, the clumps with extreme values of $\fmmc$ and CFE are simply sampled from the same underlying distribution. The fact that we do not see any clear relationship in Fig. \ref{fig:component_maps} between the locations of extrema in the velocity gradients in our \mwydyn\ fits and the positions, or extrema in the total linewidth of the identified cores, also supports this.

We see tentative evidence for two other aspects in Fig. \ref{fig:core_masses} panels e) and f). The first of these aspects is a tendency for the most IR-dark clumps to have the higher values of CFE, which is consistent with \citet{Anderson+21} who also found that clumps at different evolutionary stages (as traced by 8\,\mic\ brightness) tended to show different values of $\fmmc$, with IR-dark sources having high values, and IR-bright sources having lower values. However, the six IR-dark HFSs of that study are also represented in our sample, and so the results are not independent, and our sample size prevents us from drawing a statistically significant result on this point. If the $\fmmc$ and CFE distributions are time variable, then we need a dramatically larger sample to identify this.

Secondly, we point out that none of the clumps with $M < 700$\,M$_\odot$ -- corresponding to a surface density of around $\sigpc < 0.2$\,g\,cm$^{-2}$ -- achieve $\fmmc > 10\%$. Given the sample size, this is not hugely surprising, but the probability of drawing four consecutive samples that are below 10 per cent is somewhat unlikely, with odds of approximately 1/16. Various studies have proposed thresholds in gas density, above which star formation becomes more efficient. For example, \citet{Lada+10} proposed a threshold of 116\,\Msun\,pc$^{-2}$, above which the surface density of star formation correlates linearly with gas surface density, and \citet{Heiderman+10} proposed a similar threshold of 129\,\Msun\,pc$^{-2}$ (both values correspond to $\sim 0.025$\,g\,cm$^{-2}$). On clump scales, \citet{Traficante+20} found that a sample of 70-\mic-dark clumps that become increasingly dominated by non-thermal motions at values of surface density greater than 0.1\,g\,cm$^{-2}$. It is possible that what we see here is evidence of a similar threshold, though we note that there are still several clumps with $\fmmc < 10\%$ that exceed the threshold. If such a threshold is present, it is not convincing from these data alone, and we require a much larger sample to reach a robust conclusion.

\subsection{The role of IR-dark hub-filament systems}

\citet{Peretto+22} showed that, on average, clumps hosting HFSs are both more infrared-bright, and more massive than non-HFS clumps. Furthermore, \citet{Rigby+21} showed evidence for the accretion of material onto clumps throughout their early- to mid-stages of evolution. In this study, we have also shown that more massive clumps are associated with more massive MMCs (and core populations overall), and greater kinematic complexity in dense gas. All of these results would seem to suggest that high-convergence hub-filament systems are the \emph{result} of the evolution of clumps, as opposed to a coincidental set of initial conditions that give rise to the growth of the densest and most massive clumps. What, then, is the role of infrared-dark hub-systems, such as those of \citet{Anderson+21}? We have identified one source within our sample, SDC24.489, that bucks the trend in Fig. \ref{fig:MMCfrac}, with an extremely large value of $\qmmc$ given its $\fc$ value, compared to the rest of the sample. It is tempting to suggest that, as an extremely strong and infrared-dark HFS, SDC24.489 is something of a special source.

However, there are some observational effects that we can expect are having an impact on our interpretation of hub-filament systems, and the use of the convergence parameter $\fc$. Firstly, much of the focus on HFSs in recent years has depended on targets being identified from within \emph{Spitzer} 8-\mic\ data as infrared-dark features \citep[e.g.][]{Peretto+Fuller09, Peretto+13}, which may result in a selection bias, because we can not locate infrared-bright HFSs at similar resolution in blind surveys. In these data, the extinction features allow a view of the gas column density at a resolution of 2 arcseconds, but only where the location and status of the cloud are, in some sense, `lucky'. Sources which are located at greater distances will be more difficult to see against the diffuse background, as will those which are more evolved (with warmer gas), and we do not have any way of surveying total column density across large areas of the Galaxy at comparable resolution for sources which do not satisfy these selection criteria. Even with the \emph{Herschel} 250-\mic\ data, the resolution is a factor of \emph{nine} worse than \emph{Spitzer}, limiting the visibility of filaments to sources closer than 5\,kpc in \citet{Kumar+20}. If the 50-m AtLAST telescope \citep{Klaassen+20} is ever built, and equipped with a Band 10 (350-\mic) continuum receiver, only then will we have access to large-area and unbiased surveys of column density at these resolutions. For now, the best available data for tracking column density across a wide range of evolutionary stages of clumps appear to be the GASTON-GPS data, but even here the resolution is a factor of six worse than \emph{Spitzer}. The combination of greater resolution and higher sensitivity may partly explain why evolutionary trends in clump properties had been identified in GASTON-GPS \citep{Rigby+21} and especially with respect to filament convergence \citep{Peretto+22} that were not present in ATLASGAL \citep[e.g.][]{Urquhart+18}, though source extraction techniques also differ substantially.

Secondly, the viewing angle surely has an impact on how we study HFSs. If the filaments in HFSs are isotropically arranged in three-dimensions, then it should be expected that some viewing angles should be more favourable than others for their identification. For instance, in a HFS composed of three filaments converging on a central clump with opening angles that are as widely separated as possible, we should expect from some viewing angles to see only two of these three filaments, and we would regard it as a weaker HFS than if we could see all three filaments (and record a lower $\fc$ value). In the extreme case where HFSs may be completely planar (with some indications of this seen in \citealt{Trevino-Morales+19} and Anderson et al. in prep), we may only see their true extent -- and measure the highest $\fc$ value -- if our viewing angle is such that they are seen face-on. In this case, some fraction of IRDCs that appear as filamentary may turn out to be HFSs at a high angle of inclination. In this case, SDC24.489 may not be an outlier in Fig. \ref{fig:MMCfrac} at all, but merely have the most fortuitous viewing angle.

To unravel whether SDC24.489 really is an intrinsically extreme source we require an expanded sample that has not been selected from a sample of IRDCs.

\section{Summary and conclusions} \label{sec:conclusions}

We have examined the central $\sim$1\,pc of a sample of seven IRDCs with spectral line observations in the 3\,mm band using NOEMA and the \emph{IRAM} 30-m telescope, and an expanded sample including a further six IRDCs observed by ALMA in 3\,mm continuum. We have developed a Python program called \mwydyn, which can decompose \nthp\ (1--0) spectra (and, in principle, any molecular transition with hyperfine structure) into up to three distinct velocity components in a fully-automated manner, and is a powerful tool for kinematic analysis.

We explore the relationship between the properties of the core populations, and the kinematics traced primarily by \nthp (1--0), and several key properties of their host 1-pc-scale environments which may reveal clues about their formation; the clump mass measured within a 1-pc aperture (equivalent to the local surface density), the infrared bright fraction ($\firb$) that traces evolution, the filament convergence parameter ($\fc$) that quantifies the local morphology, and the virial parameter ($\avir$), tracing the balance of kinetic and gravitational energy. For the ALMA sources there are no GASTON-GPS-like single-dish data, and so we could only compute $\fc$ values for the sources observed with NOEMA. 

Our main conclusions are as follows:
\begin{enumerate}
    \item We identified 47 cores within the central $\sim$1-pc of our 13 target clumps at a spatial resolution of $\sim$0.08 pc. The cores range in mass between $\sim5$ and 1300\,\Msun, with beam-deconvolved radii between $\sim$0.01 and 0.2\,pc.
    
    \item The mass of the most massive cores, and the total mass in cores are both strongly correlated with the 1-pc clump mass, and occur with an approximately log-normally distributed formation efficiency over the full mass range, with $\fmmc \approx 9\% \pm 0.35$ dex, and CFE $\approx 16\% \pm 0.25$ dex, respectively.
    
    \item We report tentative evidence that clumps are most efficient at concentrating mass into their most massive cores during the earliest and most infrared-dark phases.
    
    \item We find that the fraction of the total mass in cores that resides within the single most massive core is negatively correlated with the filament convergence parameter, implying that HFSs tend to be associated with less dominant most-massive cores than in less convergent systems.

    \item Our analysis in \ceo, \hcop, HNC, and \nthp\ (1--0) consistently reveal a picture of highly dynamic clump centres, which may give rise to the log-normal distributions of $\fmmc$ and CFE.
    
    \item The weighted mean values of measures of kinematic complexity of the dense gas traced by \nthp\ (1--0), including inter-component velocity dispersion, total linewidth, and number of velocity components, increase with 1-pc clump mass and $\fc$.

    \item The distributions of fitted \nthp\ (1--0) linewidths in spectra associated with the most massive cores are qualitatively similar to the distributions for spectra associated with ambient parsec-scale material. This suggests that the MMCs inherit their kinematic properties -- at least in terms of dense gas -- from the wider (chaotic) clump-scale environment, and are not kinematically decoupled on $\sim$0.1-pc scales.
    
    \item Filamentary substructure is not found to be a general characteristic of our sample, and we instead find that apparently filament-like oscillatory structures seen in position-velocity projections arise as a consequence of projecting concave and convex structures that appear in a sheet-like morphology in 3-D (PPV)-space into 2-D (position velocity; PV)-space. This is most likely a result of observational limitations, though filament merging in the centres of the clumps may also eradicate their signatures.
    
\end{enumerate}

Our data support a picture of clump evolution in which hub-filamentary morphology is a by-product of the evolution itself, as opposed to being a result of particular initial conditions. The clump's gravitational potential draws in surrounding filamentary structures over time which, in turn, promote the growth of the MMCs at the convergence point. The MMCs grow most rapidly at early times, with an initially top-heavy core mass function. The rate of growth of the neighbouring core masses increases relative to the MMCs at later times, either as a result of accelerating infall, or MMC accretion rates slowing due to protostellar feedback, resulting in an apparent steepening of the core mass function. Stellar feedback must ultimately halt the collapse onto the hub centre, with ionising radiation from young massive stars dispersing the diffuse ambient clump material, leaving the exposed spines of the dense filaments radially arranged around the newly-formed central cluster.

We temper these results with the caveat that our sample size is small, with only between seven and thirteen sources (for the kinematic and core-based results, respectively), clump evolution is a multi-dimensional problem. A much larger sample will be returned to in the future, based on ALMA observations that are ongoing at the time of writing. What we have presented here is a powerful suite of tools -- especially \mwydyn\ -- that will become increasingly useful as sample sizes expand, enabling the community to untangle many facets of clump evolution and star cluster formation.

\section*{Acknowledgements}

We would like to thank the anonymous referee, whose careful reading and review helped to improve the quality of this manuscript. AJR would like to thank Charl\`{e}ne Lef\`{e}vre for assistance with preparing the NOEMA observations, and calibrating, imaging, and cleaning the data. AJR would like to thank the School of Physics and Astronomy at the University of Leeds for a postdoctoral fellowship. AJR and NP acknowledge the support of STFC consolidated grant number ST/N000706/1 and ST/S00033X/1. GAF also acknowledges support from the University of Cologne and its Global Faculty Program. EJW gratefully acknowledges funding from the German Research Foundation (DFG) in the form of an Emmy Noether Research Group (grant number KR4598/2-1, PI Kreckel). This work is based on observations carried out under project numbers W18AK and S19AG with the \emph{IRAM} NOEMA Interferometer and 217-18 and 097-19 with the 30m telescope. \emph{IRAM} is supported by INSU/CNRS (France), MPG (Germany) and IGN (Spain). The research leading to these results has received funding from the European Union’s Horizon 2020 research and innovation programme under grant agreement No 730562 [RadioNet].

\textbf{Software}: {\sc astropy} \citep{TheAstropyCollaboration22}, {\sc carta} \citep{Comrie+21}, {\sc lmfit} \citep{Newville+14}, {\sc matplotlib} \citep{Hunter07}, {\sc numpy} \citep{Harris+20}, {\sc scipy} \citep{Virtanen+20}, {\sc sklearn} \citep{Pedregosa+11}.

\section*{Data Availability}

A combined version of Tables \ref{tab:targets}, \ref{tab:avir}, and \ref{tab:MMcores}, Table \ref{tab:cores}, are available in machine-readable format as online supplementary material to this article. The NOEMA continuum and spectral data are publicly available on Zenodo, in a repository accessible at \url{https://doi.org/10.5281/zenodo.10455794}. The \emph{Spitzer} and \emph{Herschel} data are publicly available. The NIKA maps of SDC18.888 and SDC24.489 were published in \citet{Rigby+18}. In keeping with the \emph{IRAM} Large Program Policy\footnote{\url{https://iram-institute.org/science-portal/proposals/lp/policy/}}, the GASTON-GPS data will be made publicly available no later than October 2024, and a data release paper is being prepared (Zhou et al., in prep.). 


\bibliographystyle{mnras}
\bibliography{SDCs-NOEMA}


\appendix

\section{Aperture corrections} \label{app:aperture_corrections}
The source extraction performed in Section \ref{sec:sourceextraction} used the {\sc astrodendro} implementation of the dendrogram algorithm \citep{Rosolowsky+08}. The dendrogram algorithm extracts structures based on the topology of the image as a function of contour level. There are several aspects of the methodology which should be considered when using the integrated flux densities recovered by the algorithm, and these are related to two key aspects:

\begin{enumerate}
    \item Firstly, sources are only identified within contours which have a flux density specified by the {\texttt{min\_value}} parameter which, in our case, is equal to a value of three times the rms noise value for each source. Any flux located outside of this contour will not be included in the integrated measurement. 
    \item Secondly, sources that are not perfectly isolated may have their boundaries defined at a higher contour level, and consequently any flux located outside of this contour is not included in the measurement.
\end{enumerate}

\begin{figure}
    \centering
    \includegraphics[width=0.8\linewidth]{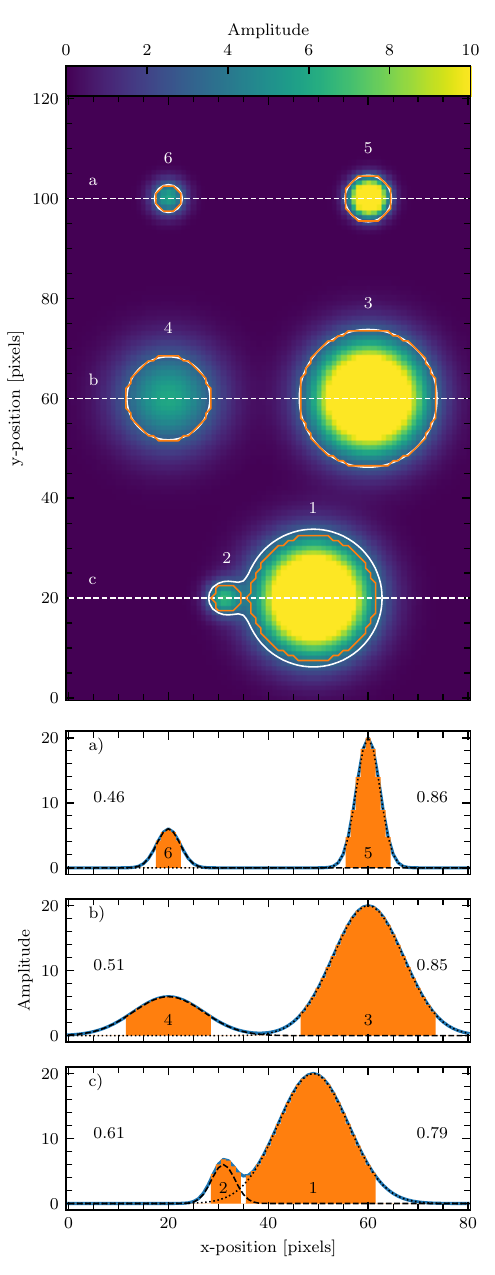}
    \caption{Case studies demonstrating the need to apply aperture corrections for compact sources extracted with {\sc astrodendro}. Image: six Gaussian sources with varying amplitudes modelled that are: compact (5 and 6), extended (1, 2, 3, 4), and partially blended (1, 2). The white contour shows the lowest level considered to be considered as part of a source (at a value of 3), while the orange contours show the boundaries of the dendrogram leaves. The bottom three panels show 1-dimensional profiles through the image at the position of the dashed white lines labelled a--c, and the shaded region shows the areas integrated for the integrated flux measurements (see text), respectively. The dashed and dotted black lines show the profile of the individual sources on the left and right of each row, respectively, while the solid blue line shows the combined profile. The numbers to the side of each model profile show the fraction of integrated flux from the model contained within the measurements on the top; the appropriate aperture correction factor is the reciprocal of this number.}
    \label{fig:apcor_profiles}
\end{figure}

In Fig. \ref{fig:apcor_profiles}, we present several case studies to illustrate why it is important to account for these effects. The image contains a total of six sources, numbered 1--6 according to the dendrogram extraction that was performed. These models present an idealised noiseless image of sources viewed by a telescope with a 3.8-pixel-FWHM Gaussian beam (similar to our observations). Sources 5 and 6 are point-like, with a peak intensity of 20 and 6, respectively, in arbitrary units. Sources 3 and 4 are extended with respect to the beam, and with peak intensities of 20 and 6, respectively. Sources 1 and 2 are extended, with peak intensities of 20 and 6, respectively, and are partially blended. The bottom three panels show profiles taken through the image at the positions of the dashed white lines, and contain shaded areas to demonstrated the intensity that is accounted for by our integrated flux density measurements. The under-reporting of integrated flux densities is worst for point sources at low signal-to-noise ratio, and the behaviour can become unpredictable in crowded environments where intensity profiles become blended; in the example here, source 2 acquires a fraction of source 1's integrated flux density.

This is perfectly appropriate for isolated sources -- although it will still under-report the flux as a consequence of point i) above -- but, in regions with either complex background emission, or for blended sources, this may incorporate unrelated flux. By contrast the clipped measurement will more severely under-report the integrated flux of isolated sources, but will do a better job of recovering the appropriate integrated flux of compact sources that are lying in top of a region of extended emission.

The 1-D profiles of the sources in Fig. \ref{fig:apcor_profiles} show the fraction of the total integrated flux (in 2-D) recovered by the clipped and bijected measurements (printed on the top and bottom, respectively`). We define the aperture correction factor, $f_\mathrm{ap}$ to be the reciprocal of these values that will, upon multiplication, restore the total flux. The models with the figure demonstrate that the integrated flux densities are more reliably recovered for sources that have a higher peak signal-to-noise ratio, and which have larger areas. A further point of note is that, by construction sources 5 and 6 represent pure point sources, with a size equal to the observation beam. The area within the contour defined by \texttt{min\_value} contains 21 pixels, compared to the nominal 36 pixels in the beam. If we had set the \texttt{min\_npix} parameter to be equal to the number of pixels in the beam, we would not have recovered this source at all, and this motivates our adoption of the half-beam-area for \texttt{min\_npix} in Section. \ref{sec:sourceextraction}.

\begin{figure}
    \centering
    \includegraphics[width=0.8\linewidth]{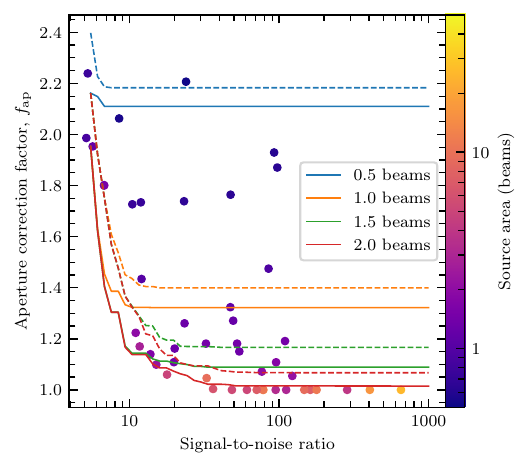}
    \caption{The aperture correction factor, $f_\mathrm{ap}$ determined as a function of peak signal-to-noise ratio (SNR) for sources with different areas with respect to the main beam size, denoted by the different colour curves. For extended and high-SNR sources, the correction factor tends to unity. These particular curves were derived to be appropriate for the synthesised NOEMA beam of SDC18.888.}
    \label{fig:apcor_factors}
\end{figure}

This far, our models have considered the simplistic case of a Gaussian telescope beam, with no noise. For our observations, we calculate a $f_\mathrm{ap}$ that is appropriate for each source, by measuring the equivalent fractions of the total flux in the integrated fluxes within the appropriate synthesised beam. The PSF of each individual observation is used, and we establish the reference flux within an elliptical aperture that gives the maximum flux up to a radius of two beam radii. For the NOEMA-observed sources, the reference aperture is typically 1.8 beam radii, while for the ALMA-observed ones, this is 2.0 beam radii, with the difference arising from the different observational setups. In Fig. \ref{fig:apcor_factors}, we show $f_\mathrm{ap}$ as a function of signal-to-noise ratio for sources with different areas ranging from 0.5 to 2 beam-areas for SDC18.888.

\section{\mwydyn} \label{app:mwydyn}

\subsection{Description of algorithm} \label{app:mwydyn_procedure}
The procedure runs as follows:
\begin{enumerate}
    \item First, the data cube is converted from units of surface brightness (Jy$\,$beam$^{-1}$) into main beam brightness temperature (K).
    
    \item A noise map is generated by calculating the rms of the first and last 25 channels  of each spectrum, which is used to guide the algorithm to only fit the spectra which have a peak signal-to-noise ratio (S/N) of greater than 10. 
    
    \item All of the valid pixels are cycled through one-by-one. First, a single \nthp (1--0) component fit is performed on the spectrum. The fitting is performed using the Levenberg-Marquardt algorithm-based \texttt{lmfit} Python package \citep{Newville+14}, which requires initial guesses of the four parameters. \texttt{lmfit} offers several advantages over the more widely used \texttt{scipy.optimize}, including providing support for constraints on the parameters.
    
    For $p_1$, we simply take the amplitude of the brightest channel as the initial guess, and this channel also provides the initial guess for the centroid velocity $p_2$. For the linewidth $p_3$ the initial guess is 0.5 \kms, which roughly corresponds to the FWHM linewidth corresponding to the isothermal sound speed at 10\,K, and for the total opacity $p_4$, we adopt a value of 0.2 on the basis that \nthp\ (1--0) is generally optically thin.

    \item Next, the algorithm attempts to fit second and third velocity components to the spectrum, comprising of 8- and 12-parameter models, respectively. The initial guesses for the first four parameters (i.e. the parameters for the first component) are made in the same way as in the previous step. For the second set of parameters, a duplicate of the first component initial guesses is used, but with a new centroid velocity guess. This new centroid velocity guess is estimated by comparing the velocity range of detected emission (i.e. the first and last channels which have intensities exceeding an S/N of 3.5), with the velocity range expected for the single-component fit (i.e. $\sim$15\,\kms, assuming that only its hyperfine multiplets were detected. The value of 3.5 for the minimum S/N for `detected' emission in this case was chosen such that a spurious noise spike would register as a false positive once in every $\sim$2000 channels, which greatly exceeds the 500 channels of the data cubes in this study.
    
    For the third component, the same initial guess values are adopted as for the second component.
    
    \item Each of the spectra are now cycled through again, in order to determine which of the single, double or triple-component models best fit the data. This is primarily done by comparing the Bayesian Information Criterion (BIC) quality of fit estimator for each of the fitted models. The BIC encapsulates the likelihood function, but includes a punishment term for the number of fit parameters such that increasing the number of fit parameters (i.e. adding additional velocity components) does not systematically result in an improved quality of fit. For each spectrum, the model resulting in the lowest BIC is selected as the preferred model, but for the higher-order components, we impose that they must constitute an improvement in BIC (i.e. a lower value) of at least 20 over the lower-order models.
    
    \item Thus far each spectrum has been treated independently of its neighbours, and while \texttt{lmfit} generally provides an excellent exploration of the parameter space, better solutions to similar spectra may have been found in the vicinity. Since the pixel size in our data cubes oversample the beam with $\sim$4--5 pixels across the major and minor axes, adjacent pixels are strongly correlated. For the final step we cycle through each spectrum in our initial fit, and identify any spectra within a search radius of 2 pixels which have better BIC, and attempt a refit using the corresponding fit parameters as the initial guesses. This process is repeated 10 times to allow any better-fitting parameters percolate across the pixel grid, though on these data, the vast majority of spectra take on 3 or fewer refits. This approach is similar to that of \citet{Koch+21}.
    
    \item With the fitting now complete, the results are written to a table containing a list of each component in each spectrum, along with its fit parameters, and \texttt{.fits} cubes of the model, and images of the ancillary data products such as BIC and component maps.
\end{enumerate}

There are a number of parameters that control the operation of \mwydyn. Firstly, the \texttt{snrlim} parameter sets the minimum peak signal-to-noise ratio for a spectrum for \mwydyn\ to attempt to perform a fit. We find that a value of 10 is optimum for this procedure, due to the relative height of the faintest hyperfine component: below this level, the so-called `isolated' component begins to reach a signal-to-noise ratio of around 3, which has negative implications for our initial guesses for the velocity centroid of the second component as detailed earlier in this Section. Secondly, \texttt{delbic} controls the improvement in the BIC of which an additional component must provide in order to be accepted as a better solution. This value is set to 20 by default, which provides a good compromise between including minor new components to fit non-Gaussian line shapes, but still having the flexibility to fit a wide range of line profiles. The \texttt{refrad} parameter controls the number of pixels in the search radius for the spatial refinement stage, and the default value of 2 allows new solutions to be found reasonably rapidly without the increased computational costs of having a higher value. For \texttt{cleaniter}, the number of iterations of spatial refinement, a value of 10 provides enough leeway for new solutions to propagate across the image region. In the case of SDC18.888, this is more than enough, for which no spectrum is re-fitted more than 4 times in this stage.

\subsection{Fitting results} \label{app:3dfigs}
\begin{figure*}
    \centering
    \includegraphics[trim={0 6mm 0.8mm 2mm}, clip, width=0.49\textwidth]{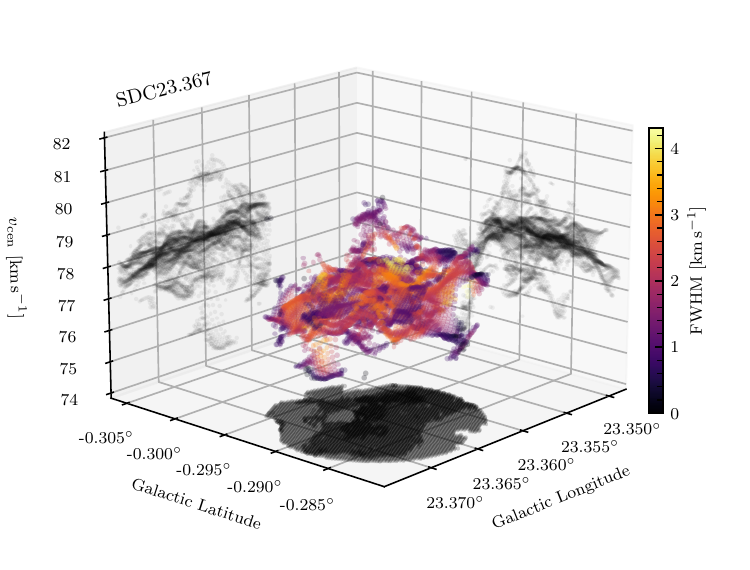}
    \includegraphics[trim={0 6mm 0.8mm 2mm}, clip, width=0.49\textwidth]{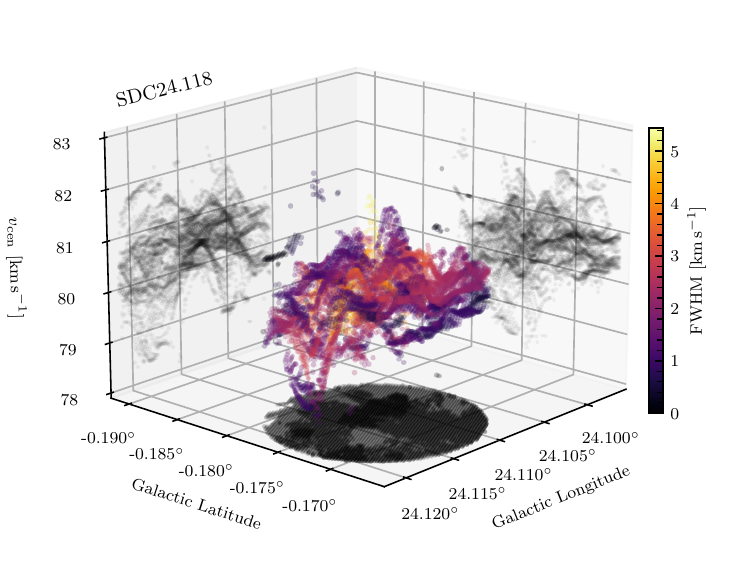}
    \includegraphics[trim={0 6mm 0.8mm 2mm}, clip, width=0.49\textwidth]{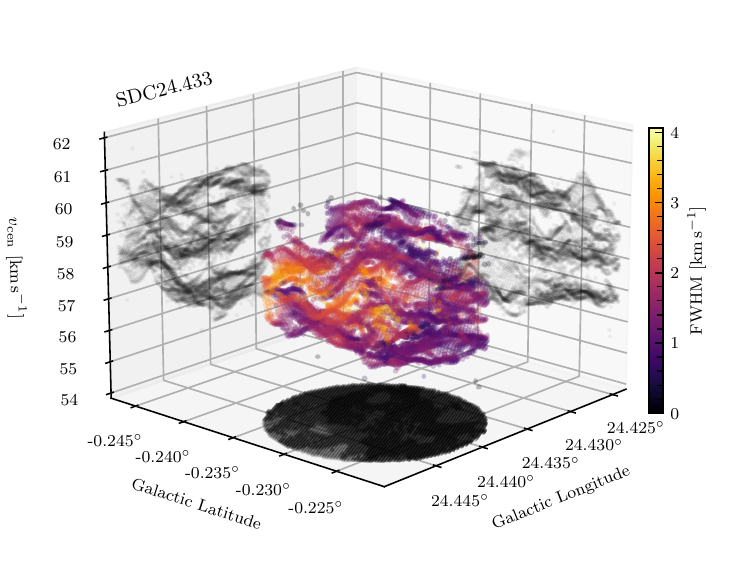}
    \includegraphics[trim={0 6mm 0.8mm 2mm}, clip, width=0.49\textwidth]{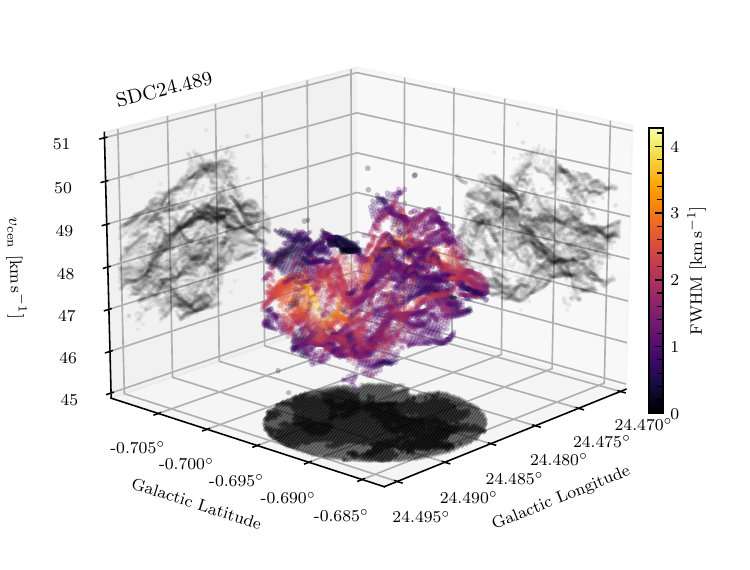}
    \includegraphics[trim={0 6mm 0.8mm 2mm}, clip, width=0.49\textwidth]{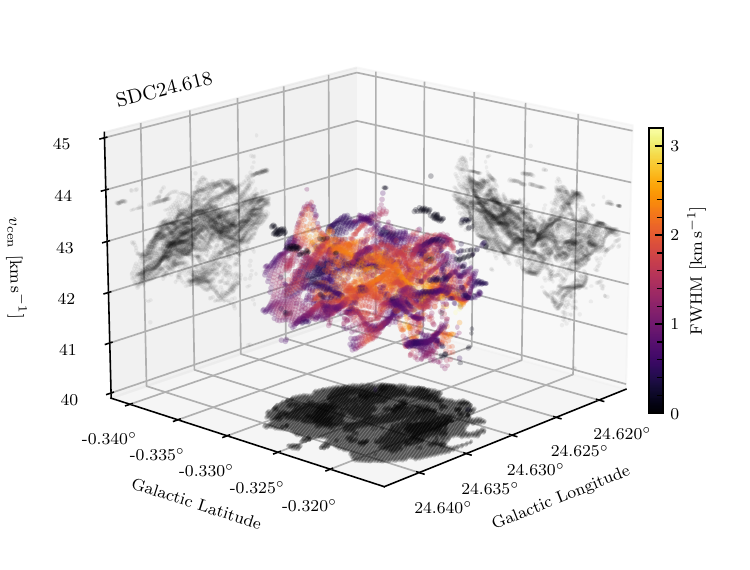}
    \includegraphics[trim={0 6mm 0.8mm 2mm}, clip, width=0.49\textwidth]{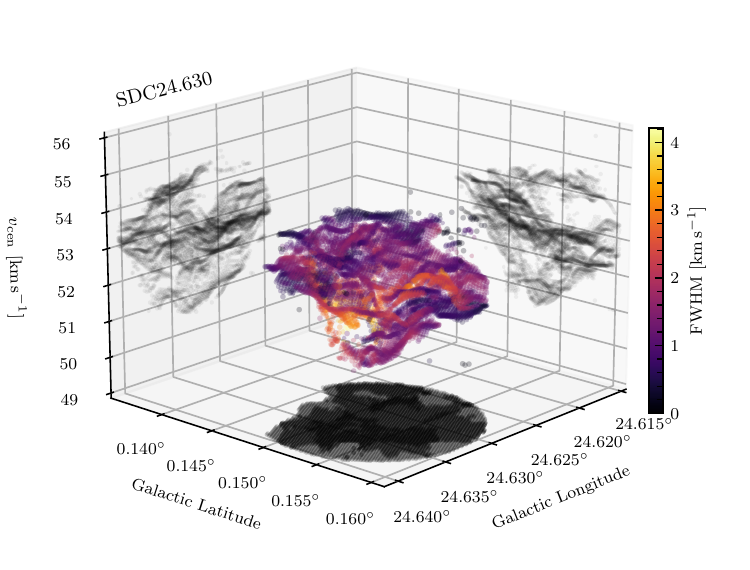}
    \caption{As in Fig. \ref{fig:mwydyn3d}, but for SDC23.367, SDC24.118, SDC24.433, SDC24.489, SDC24.618, and SDC24.630.}
    \label{fig:3dfigs}
\end{figure*}

In Sec. \ref{sec:mwydyn} we described \mwydyn, a fully-automated multiple-velocity-component fitting algorithm for hyperfine spectra, and we displayed the resulting position-position-velocity diagram for SDC18.888. In Fig. \ref{fig:3dfigs}, we show the same form of diagram with the other six sources in our NOEMA sample.

\section{Component maps} \label{app:componentmaps}

\begin{figure*}
    \centering
    \includegraphics[width=\linewidth]{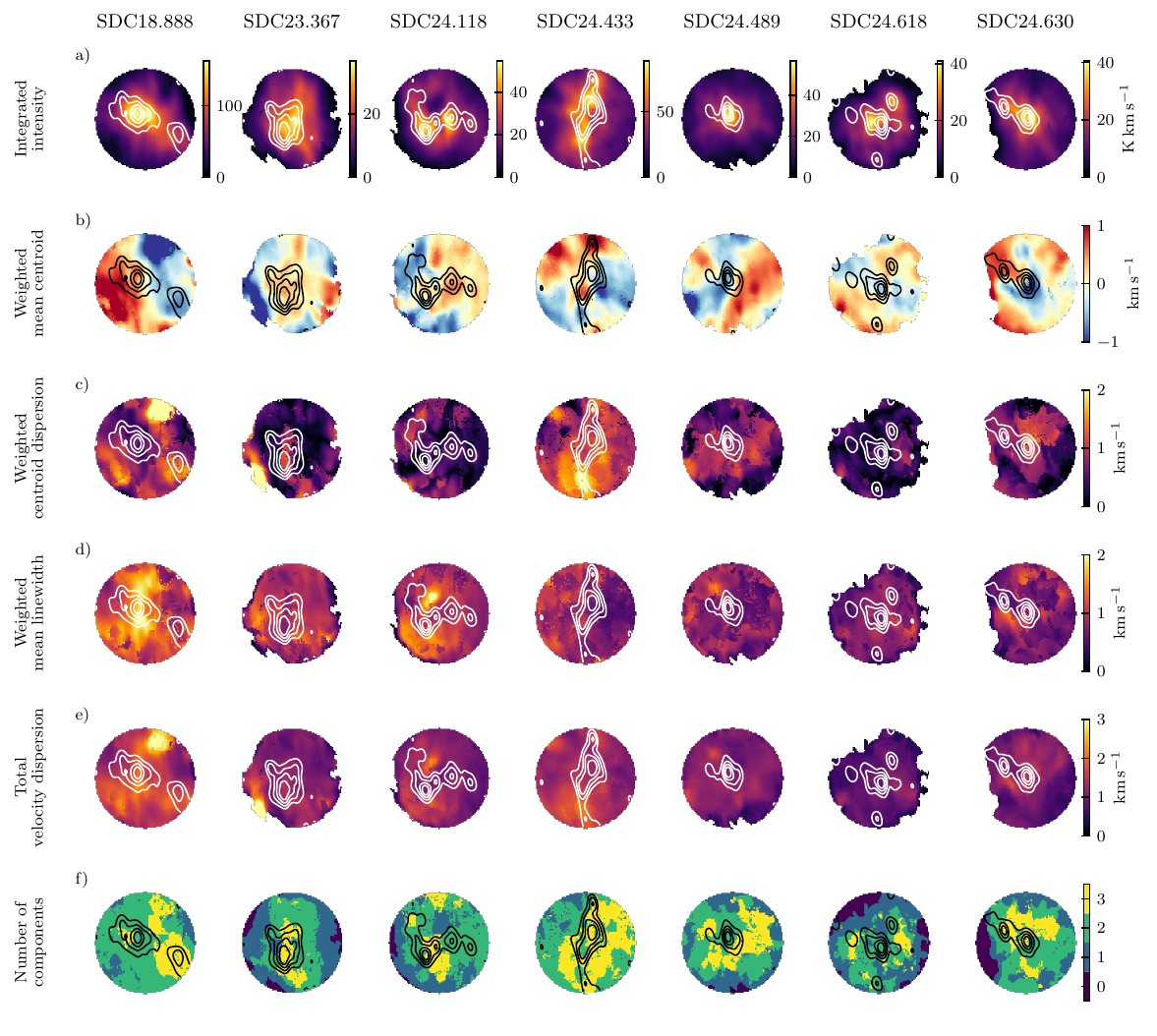}
    \caption{Various statistical quantities derived from the \mwydyn\ fitting results to each of the seven IRDCs observed with NOEMA. Each column corresponds to a particular target, and the rows are as follows: a) Integrated intensity of all components; b) Integrated intensity-weighted mean centroid velocity; c) Integrated intensity-weighted centroid dispersion with respect to the systemic velocity; d) Integrated intensity-weighted mean linewidth (i.e. FWHM / $\sqrt{8 \ln(2)}$; e) Total velocity dispersion; f) Number of fitted components per spectrum.}
    \label{fig:component_maps}
\end{figure*}

In Sec. \ref{sec:component_statistics}, we described several statistics, $\sigcomp$, $\sigline$, and $\sigcomp$, that measure the overall level of kinematic complexity in each clump by considering the ensemble of \mwydyn\ components as a single distribution. An alternative way to examine these properties is on a per-spectrum basis, which allows the construction of maps. In Fig. \ref{fig:component_maps} we display these maps and overlay the 3.2\,mm continuum contours to illustrate where the dust-traced dense gas resides. No consistent pattern between any of these properties and the location of the brightest continuum structures are seen, with the exception of the integrated intensity, which shows a reasonably strong coincidence with the continuum emission. Some of the cores (e.g. SDC18.888, SDC24.489, and SDC24.630) appear to be located in regions where the weighted mean centroid maps have a steep gradient, but this is not always the case. Some of the cores also seem to be located at local minima in the maps of mean linewidth (e.g. SDC23.367, SDC24.118, SDC24.630), but SDC18.888 presents a counter-example. The most massive core in SDC18.888 is associated with a local minimum in the map of the weighted centroid dispersion, but has a high value in the weighted mean linewidth, which suggests that the core resides at a convergence point of components which have large linewidths.

\bsp	
\label{lastpage}
\end{document}